\documentclass[aps,reprint,twocolumn,longbibliography,superscriptaddress,notitlepage,footinbib]{revtex4-1}
\usepackage{amsmath,amssymb,amsfonts}
\usepackage[pdftex]{graphicx}
\usepackage{bm}
\usepackage{color}
\usepackage{xcolor}
\usepackage{float}
\usepackage{braket}
\usepackage[colorlinks=True,linkcolor=blue,citecolor=blue,urlcolor=blue]{hyperref}
\usepackage[all]{hypcap}
\newcommand{\pd}{{\phantom\dagger}}
\newcommand{\nc}{N_\mathcal{C}}
\newcommand{\nh}{N_\mathcal{H}}

\newcommand{\comment}[1]{}

\begin{document}
\title{Percolation in Fock space as a proxy for many-body localisation}

\author{Sthitadhi Roy}
\email{sthitadhi.roy@chem.ox.ac.uk}
\affiliation{Physical and Theoretical Chemistry, Oxford University, South Parks Road, Oxford OX1 3QZ, United Kingdom}
\affiliation{Rudolf Peierls Centre for Theoretical Physics, Clarendon Laboratory, Oxford University, Parks Road, Oxford OX1 3PU, United Kingdom}

\author{J.~T.~Chalker}
\email{john.chalker@physics.ox.ac.uk}
\affiliation{Rudolf Peierls Centre for Theoretical Physics, Clarendon Laboratory, Oxford University, Parks Road, Oxford OX1 3PU, United Kingdom}

\author{David E. Logan}
\email{david.logan@chem.ox.ac.uk}
\affiliation{Physical and Theoretical Chemistry, Oxford University, South Parks Road, Oxford OX1 3QZ, United Kingdom}

\begin{abstract}
We study classical percolation models in Fock space as proxies for the quantum many-body localisation (MBL) transition. 
Percolation rules are defined for two models of disordered quantum spin-chains using their microscopic quantum Hamiltonians and the topologies of the associated Fock-space graphs. The percolation transition is revealed by the statistics of Fock-space cluster sizes, obtained by exact enumeration for finite-sized systems. As a function of disorder strength, the typical cluster size shows a transition from a volume law in Fock space to sub-volume law, directly analogous to the behaviour of eigenstate participation entropies across the MBL transition. Finite-size scaling analyses for several diagnostics of cluster size statistics yield mutually consistent critical properties. We show further that local observables averaged over Fock-space clusters also carry signatures of the transition, with their behaviour across it in direct analogy to that of corresponding eigenstate expectation values across the MBL transition. The Fock-space clusters can be explored under a mapping to kinetically constrained models. Dynamics within this framework likewise show the ergodicity-breaking transition via Monte Carlo averaged local observables, and yield critical properties consistent with those obtained from both exact cluster enumeration and analytic results derived in our recent work~\href{https://arxiv.org/abs/1812.05115}{[arXiv:1812.05115]}. This  mapping allows access to system sizes two orders of magnitude larger than those accessible in exact enumerations. Simple physical pictures based on freezing of local real-space segments of spins are also presented, and shown to give values for the critical disorder strength and correlation length exponent $\nu$ consistent with numerical studies.
\end{abstract}

\maketitle

\section{Introduction \label{sec:intro}}

The study of quantum phase transitions~\cite{sachdev2011quantum} has formed one of the cornerstones of modern condensed matter physics, which is founded partly on the broad problem of classifying phases of matter.
Historically, much of the effort in this direction has been devoted to understanding critical phenomena hosted by ground states of many-body quantum systems. Along the way, seminal ideas such as the renormalisation group~\cite{fisher1974renormalization,wilson1975renormalization} and quantum-to-classical mappings~\cite{suzuki1976relationship} have 
been developed and applied with immense success.
However, frameworks for full characterisation of generic many-body quantum systems via all the eigenstates of the Hamiltonian governing the system continue to elude us. 
This issue has recently gained 
prominence, as it has been realised that the notion of quantum criticality is not just limited to ground states, but extends to arbitrary excited eigenstates with finite energy 
densities~\cite{pal2010many,huse2013localisation,pekker2014Hilbert,paramesawran2018many} and even to out-of-equilibrium 
systems~\cite{khemani2016phase,moessner2017equilibration,roy2018dynamical,roy2018nonequilibrium,chan2018spectral,berdanier2018floquet}.
At the heart of 
much of 
this lies the physics of many-body localisation, where eigenstates at arbitrary energy densities of disordered interacting quantum systems undergo a localisation transition at a critical value of the disorder strength which may depend on the energy density~\cite{gornyi2005interacting,basko2006metal,oganesyan2007localisation,znidaric2008many,pal2010many,kjall2014many,laumann2014many,luitz2015many,lev2015absence,baldwin2016manybody} (see Refs.~\cite{nandkishore2015many,abanin2017recent,alet2018many} for reviews).

Some attempts at understanding universal properties of the many-body localisation transition have involved treating interaction-induced resonances hierarchically, in a renormalisation group-like procedure within a phenomenological coarse-grained model~\cite{potter2015universal,vosk2015theory,dumitrescu2017scaling,goremykina2018analytically,dumitrescu2018kosterlitz}.
While the specifics of the RG scheme vary among these works, a common feature is that the empirical 
criterion for resonance maps the system to a classical model, thus allowing much larger system sizes computationally, or even analytical solutions.

A complementary approach was introduced recently by us, in which a classical percolation transition in the Fock space of a disordered quantum system was shown to capture certain aspects of the many-body localisation transition~\cite{roy2018exact}.
The approach relies on the fact that the Hamiltonian of a quantum system, in general, can be written as a tight-binding Hamiltonian in Fock space under a choice of basis states, and the hoppings therein represent possible many-body resonances.
By using a classical criterion for the resonance to 
occur or not, an edge between a pair of basis states on the Fock-space graph is defined to be present (`active') or absent. This in turns defines the percolation problem. 
We remark that, from a purely quantum mechanical point of view, a self-consistent mean-field approach based on the Fock-space tight-binding model was recently shown to capture aspects of the many-body localisation 
problem~\cite{logan2018many}.

In our earlier work~\cite{roy2018exact}, the Fock-space percolation problem was introduced using a disordered tilted-field Ising (TFI) model. While an exact solution was obtained for the critical disorder and the  
correlation length exponent, this work also raised 
a number of significant questions that invite further exploration:
\begin{itemize}
	\item Is the percolation transition specific to the particular microscopics of the TFI model, or is it more general? 
	\item Can one gain physical insight into the nature of the microscopic processes that dominate the physics near the transition?
	\item Does detailed numerical study of the statistics of cluster sizes yield critical properties consistent with the analytical solution?
    \item Can one expose analogies between the behaviour of local (real space) observables across the percolation transition, and the quantum case?
    \item Can one exploit a possible mapping between the percolation problem and kinetically constrained dynamics, such that Monte Carlo dynamics can be used  to extract critical properties, but for much larger system sizes?
\end{itemize}
In this work we seek to answer these questions. Sec.~\ref{sec:overview} provides an overview of the paper.

\subsection{Overview \label{sec:overview}}

\begin{figure}
\includegraphics[width=\columnwidth]{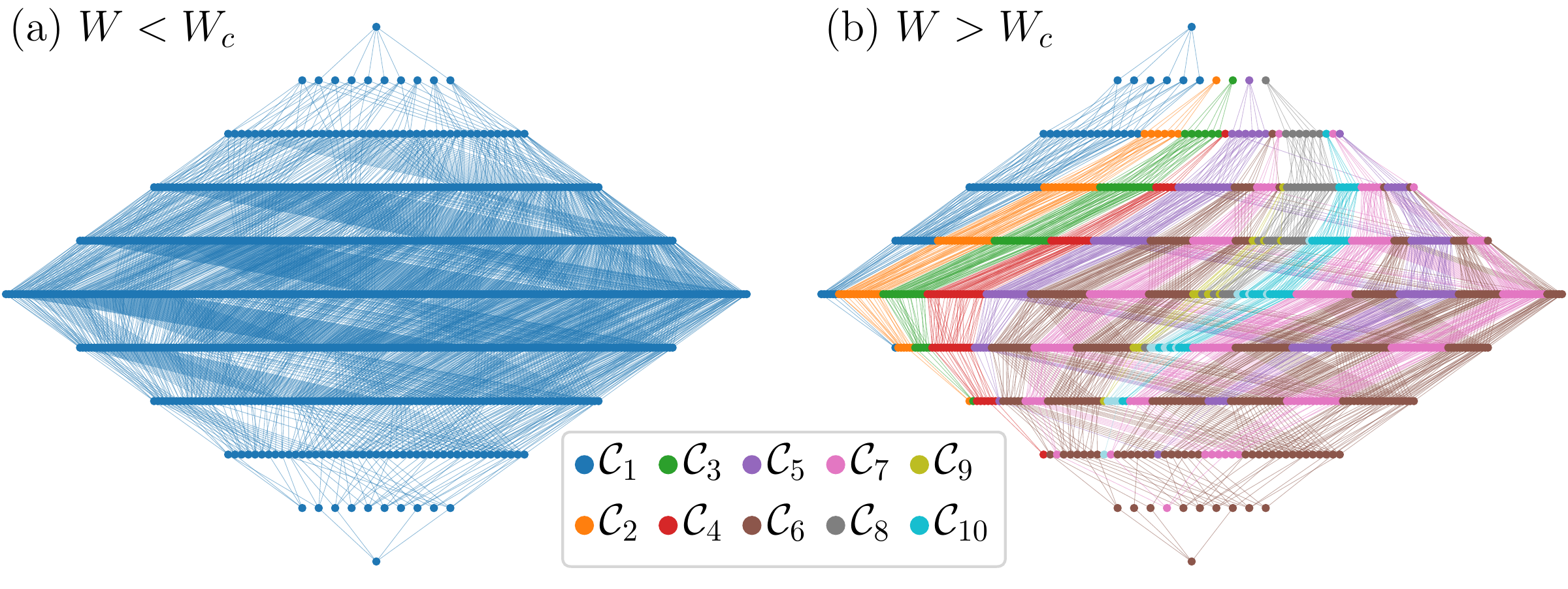}
\caption{An overview of the percolating delocalised phase (a) and the non-percolating localised phase (b) on the Fock-space graph. The graph shown is the Fock space of a disordered Ising model [Eq.~\eqref{eq:hamising}] with ten spins, where each node represents a product state in the $\sigma^z_\ell=\pm1$ basis. The different colours show disjoint clusters subject to the percolation criterion 
described in Sec.~\ref{sec:percolationrules}. In the percolating phase, all nodes belong to the same cluster as indicated by the same colour throughout the graph. By contrast, in the non-percolating phase the Fock space splits up into many small clusters as indicated by different colours.
In the particular disorder realisation shown for the non-percolating phase in (b), there are 10 clusters, labelled as $\mathcal{C}_1$ through $\mathcal{C}_{10}$, with the corresponding colours indicated in the legend.}
\label{fig:network}
\end{figure}

The overall picture of the two phases emerging from the present work is summarised in Fig.~\ref{fig:network}. This shows the Fock-space graph for the disordered Ising (TFI) chain in a product state basis, along with classical percolation clusters derived from the mapping we introduce.
All nodes of the graph shown with the same colour belong to the same cluster.
In the delocalised phase, all nodes have the same colour and hence belong to a single cluster, which percolates in the sense that its size is proportional (equal in this case) to the Fock-space dimension.
On the other hand, in the localised phase, Fock space fragments into many clusters (a diverging number in the thermodynamic limit), each of which has a size that is a vanishing fraction of the Fock-space dimension.
This is indicated graphically in Fig.~\ref{fig:network} by many distinct clusters of different colours.
It is also important to note that, due to the thermodynamically large local connectivity of the Fock-space graph, 
 our percolation problem is very different from 
standard percolation problems. This is readily seen in the nature of the clusters in Fock space in the localised phase; 
a given cluster can 
straddle the Fock space yet contain only a vanishing fraction of the Fock-space sites.

We begin in Sec.~\ref{sec:percolationmodel} by describing the percolation problem on Fock space.
Sec.~\ref{sec:physical} then presents a detailed physical picture of the two phases and of the transition. In the presence of interactions, we show that there exists a non-trivial percolating phase where all the nodes of the Fock-space graph lie in the same cluster 
but not all edges on the graph are active, this being the defining feature of the percolating phase close to the phase transition.
We further argue that the transition is driven by freezing of spin-configurations of finite length segments on the chain, which is the characteristic feature of the non-percolating phase in the vicinity of the transition. This picture demonstrates the key role played at the classical transition by spin-spin interactions mapped from the quantum Hamiltonian.
Interestingly, we find that the dominant microscopic processes near the critical point are rather different between the TFI and XXZ models, yet the critical exponents that we 
deduce are the same, indicating their universality.

In Sec.~\ref{sec:numerical} we consider the disordered XXZ chain, studying a variety of numerical diagnostics for the transition and extracting the critical disorder strength and exponents. 
In particular, we look at the probability distribution of cluster sizes, and their average and typical sizes. 
We find that they do indeed act as diagnostics of the phase transition, 
and we perform a finite-size scaling analysis to extract the critical disorder and exponent. 
Fluctuations of the cluster sizes also show a peak at the transition, characteristic of susceptibilities.
While the percolation problem is set up in Fock space, an important result is that local (real-space) observables, such as magnetisations, appropriately defined over the cluster, also act as diagnostics of the phase transition.
All such numerical diagnostics yield values of the critical disorder and the correlation length scaling exponent consistent with each other.  This constitutes the first of the two main results of this work.
The numerical results also show clear parallels to the MBL transition in quantum systems. For example,
the typical cluster size is shown to be directly analogous to the participation entropies of quantum eigenstates, and their scaling with Fock-space dimension across the transition 
is akin to that of participation entropies. 
Likewise, the cluster-averaged local magnetisation distributions show a transition analogous to that displayed by their eigenstate expectation values in the quantum case.

The second main result of this work, contained in Sec.~\ref{sec:montecarlo}, is that our classical percolation model can be interpreted as an instance of kinetically constrained models~\cite{ritort2003glassy,garrahan2011kinetically}.
This is particularly interesting as such models were historically developed as models for glass formers~\cite{fredrickson1984kinetic,fredrickson1985facilitated}, and hence host ergodicity-breaking phase transitions. 
In fact, quantum models inspired from kinetically constrained models have also been shown to exhibit quasi-many body localisation in the absence of disorder~\cite{horssen2015dynamics,hickey2016signatures,lan2018quantum}.
Our mapping to such a kinetically constrained model allows us to study dynamics under a suitably defined Monte Carlo scheme, allowing access to system sizes some two orders of magnitude larger than those accessible in the exact enumeration of the Fock space.
We show that the Monte Carlo history of appropriately defined local observables reveals the phase transition, and critical disorder and exponents extracted in this way
are found to be consistent with those obtained in Sec.~\ref{sec:numerical}.

Finally, we close with concluding remarks and an outlook in Sec.~\ref{sec:conclusion}.

\section{Percolation in Fock space \label{sec:percolationmodel}}

In this section 
we formulate a classical bond percolation problem in the Fock space of a 
quantum many-body system.  
We view the Hamiltonian as a tight-binding model in
Fock space~\cite{welsh2018simple,logan2018many} and translate this to a classical problem by replacing off-diagonal matrix elements with bonds that are present or absent, according to an empirical but physically motivated criterion.

\subsection{Percolation rules \label{sec:percolationrules}}

Let $\{\ket{I}\}$ denote a set of many-body basis states, which we take to be nodes of a graph. The Hamiltonian of a quantum system is a tight-binding model on this graph
with the form
\begin{equation}
\mathcal{H} = \sum_I \mathcal{E}_I \ket{I}\bra{I} + \sum_{I\neq K}T_{IK}\ket{I}\bra{K}\,.
\label{eq:tightbinding}
\end{equation}
The $\mathcal{E}_I$'s are on-site energies on the Fock-space graph.
Off-diagonal matrix elements $T_{IK}\neq0$ generate hopping between nodes $I$ and $K$, and are represented by edges of the graph.
The extent of hybridisation between states $\ket{I}$ and $\ket{K}$, with $\vert\mathcal{E}_I-\mathcal{E}_K\vert = \Delta$ and 
$T_{IK}=J$, 
is proportional to $J/\sqrt{\Delta^2+J^2}$, so that
if $\Delta\gg J$ the two states are not resonant, while if $\Delta\ll J$ they are.

We translate this resonance criterion to a percolation rule on the Fock-space as follows.
Classically, an edge on the graph between two nodes $\ket{I}$ and $\ket{K}$ is \textit{active} if
\begin{equation}
\vert\mathcal{E}_I-\mathcal{E}_K\vert < \vert T_{IK}\vert.
\label{eq:active}
\end{equation}
Two nodes $\ket{I_i}$ and $\ket{I_f}$ are in the same cluster 
(even if $T_{I_iI_f}=0$) if they are connected by a continuous sequence of {active} edges.

\subsection{Models}

In this paper we consider two classical Fock-space percolation problems, derived from two disordered quantum spin chains which have been used extensively in studies of many-body localisation: 
a disordered TFI chain~\cite{imbrie2016many},
and a disordered XXZ model~\cite{oganesyan2007localisation,znidaric2008many,pal2010many,luitz2015many,lev2015absence}.
The quantum Hamiltonian for the TFI  model is
\begin{equation}
\mathcal{H}_\mathrm{TFI}^\pd = J_z^\pd\sum_{\ell=1}^{N-1}\sigma^z_\ell\sigma^z_{\ell+1} + \sum_{\ell=1}^Nh_\ell^{\pd}\sigma^z_\ell + J\sum_{\ell=1}^N\sigma^x_\ell\,.
\label{eq:hamising}
\end{equation}
For the XXZ model it is
\begin{equation}
\mathcal{H}_\mathrm{XXZ} = \sum_{\ell=1}^{N-1}[J(\sigma^x_\ell\sigma^x_{\ell+1}+\sigma^y_\ell\sigma^y_{\ell+1})+J_z^\pd\sigma^z_\ell\sigma^z_{\ell+1}]+\sum_{\ell=1}^N h_\ell^\pd\sigma^z_\ell\,.
\label{eq:hamxxz}
\end{equation}
In both cases $h_\ell\in[-W,W]$ are random fields, drawn independently at each site from a uniform distribution,
and we take $J, J_{z} >0$. 

We take the number of real-space sites in the system to be $N$, and the size of the Hilbert space to be $N_\mathcal{H}$.
We choose product states $\{\ket{I}\} \equiv \{|\{\sigma_{l}^z\}\rangle \}$ of eigenvectors of $\sigma_l^z$ as our Fock-space basis. These are exact eigenstates of the quantum Hamiltonian in the strong disorder limit, and in this limit the classical percolation system consists of isolated Fock-space sites. 
The Fock-space site energies 
$\mathcal{E}_I = \langle I \vert J_z^\pd\sum_{\ell=1}^{N-1}\sigma^z_\ell\sigma^z_{\ell+1}+\sum_{\ell=1}^Nh_\ell^{\pd}\sigma^z_\ell\vert I\rangle$ 
are straightforward to evaluate.

With this choice of basis, the Fock space graph 
for the TFI model is 
an 
$N$-dimensional hypercube with $N_\mathcal{H}=2^N$ nodes and $N_\mathcal{H}N/2$ edges, each corresponding to a single 
spin flip.
Formally, these hopping matrix elements can be represented as 
\begin{equation}
T_{IK}=\bra{I}J\sum_{\ell=1}^{N}\sigma^x_\ell\ket{K},
\label{eq:TIKtfi}
\end{equation}
where the $N$ terms in Eq.~\eqref{eq:TIKtfi} correspond to the $N$ possible spin-flips from the node $K$.

For the XXZ model, the hopping matrix elements on the Fock-space graph generate exchange of a pair of nearest-neighbour anti-parallel spins, with
\begin{equation}
T_{IK} = \bra{I}J\sum_{\ell=1}^{N-1}[\sigma^x_\ell\sigma^x_{\ell+1}+\sigma^y_\ell\sigma^y_{\ell+1}]\ket{K}\,.
\label{eq:TIKxxz}
\end{equation}
In this case total magnetisation $\sum_{\ell=1}^N\sigma^z_\ell$ is conserved and  
the graph is not a perfect hypercube~\cite{welsh2018simple}. We work in the sector with zero total magnetisation.

\section{Physical picture for the phases \label{sec:physical}}

In this section we explain why the classical models introduced in Sec.~\ref{sec:percolationmodel} 
must have two phases, percolating and localised, and hence a phase transition.
We also provide a physical picture for the nature of the two phases and discuss why the spin-spin interactions in the quantum Hamiltonians play a key role in the critical phenomena hosted by the classical models.
We present the bulk of our discussion in Sec.~\ref{sec:physicaltfi} using the TFI model as an example, and summarise equivalent results for the XXZ model in Sec.~\ref{sec:physicalxxz}.

\subsection{TFI spin chain \label{sec:physicaltfi}}

For the TFI spin chain Eq.\ \eqref{eq:hamising}, the energy cost $\Delta$ 
of flipping spin $\ell$ from $\uparrow$ to $\downarrow$ depends on the spin 
orientations of its nearest-neighbour sites $\ell\pm 1$, and is given by
\begin{align}
\vert\cdots\uparrow\uparrow\uparrow\cdots\rangle\leftrightarrow\vert\cdots\uparrow\downarrow\uparrow\cdots\rangle\Rightarrow& \Delta_\ell^{(p\uparrow)} = \vert 2h_\ell^\pd+4J_z^\pd\vert, \nonumber\\
\vert\cdots\downarrow\uparrow\downarrow\cdots\rangle\leftrightarrow\vert\cdots\downarrow\downarrow\downarrow\cdots\rangle\Rightarrow& \Delta_\ell^{(p\downarrow)} = \vert 2h_\ell^\pd-4J_z^\pd\vert, \nonumber\\
\vert\cdots\uparrow\uparrow\downarrow\cdots\rangle\leftrightarrow\vert\cdots\uparrow\downarrow\downarrow\cdots\rangle\Rightarrow& \Delta_\ell^{(a)} = \vert 2h_\ell^\pd\vert .
\label{eq:deltaap}
\end{align}
The existence of three 
energy scales, $\Delta_\ell^{(p\uparrow)}$, $\Delta_\ell^{(p\downarrow)}$ and $\Delta_\ell^{(a)}$, 
implies that for a given disorder configuration there are fractions $f_a$ and $f_n$ of spins that are respectively always and never flippable (regardless of the configuration of the neighbours), and a fraction $f_s=1-f_a-f_n$ that are flippable only for some configurations of their neighbours.

If 
$f_n>0$, 
the system is in a localised phase.
The condition for a spin at site $\ell$ never to be flippable is
\begin{equation}
\Delta_\ell^\mathrm{(min)} = \mathrm{min}\{\Delta_\ell^{(p\uparrow)},\Delta_\ell^{(p\downarrow)},\Delta_\ell^{(a)}\} >J\,.
\label{eq:deltamin}
\end{equation}
When this holds,
none of the edges involving flipping of this spin
is active. 
Hence there are at least two clusters: on one, all nodes have $\sigma^z_\ell=1$; on the other 
all nodes have $\sigma^z_\ell=-1$.
By extension, if there exists a finite fraction $f_n$ of spins that satisfy Eq.\ \eqref{eq:deltamin} and hence can never flip, Fock space 
necessarily splits up into at least $2^{f_nN}$ clusters. Each cluster has a maximum size $2^{(1-f_n)N}$. Since this is
a vanishing fraction of the Fock-space dimension in the thermodynamic limit, 
it implies
a localised phase.

Similarly, if $f_a = 1$ the system is in a percolating phase. The condition for a spin at site $\ell$ always to be flippable is
\begin{equation}
\Delta_\ell^\mathrm{(max)} = \mathrm{max}\{\Delta_\ell^{(p\uparrow)},\Delta_\ell^{(p\downarrow)},\Delta_\ell^{(a)}\} < J.
\label{eq:deltamax}
\end{equation}
For $f_a = 1$, all edges on the hypercube satisfy the percolation criterion and so all nodes belong to a single cluster, 
indicating a trivial percolating phase.

\begin{figure}
\includegraphics[width=\columnwidth]{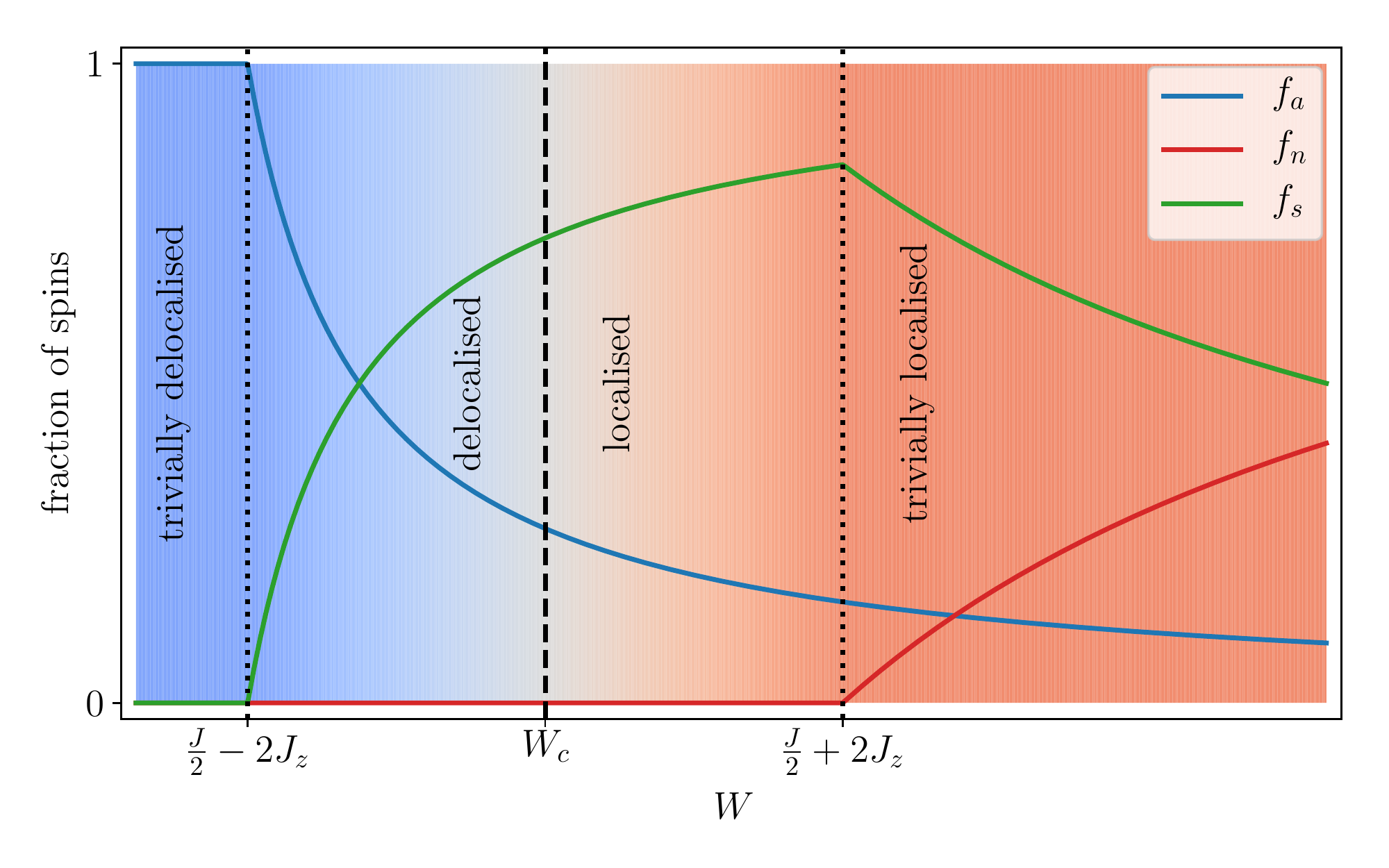}
\caption{Behaviour of the three spin fractions, $f_a$ (always flippable), $f_n$ (never flippable), and $f_s$ (flippable under some configurations of neighbours), shown as a function of disorder strength $W$,
with $W_c$ the critical disorder strength for the phase transition.
The two vertical dotted lines at $W_c^\pm = \frac{J}{2}\pm 2J_z$ denote the bounds on the critical disorder. 
For $W<W_c^-$, the system is trivially percolating, whereas for $W>W_c^+$ the system is trivially 
localised.}
\label{fig:fraction}
\end{figure}

Evaluation of $f_n$ and $f_a$ as a function of $W$ hence gives bounds on the location of the transition.
We have
\begin{subequations}
\begin{equation}
f_n = \int dh~ P(h)\Theta(\Delta^\mathrm{(min)}-J)
\end{equation}
and
\begin{equation}
f_a = \int dh~ P(h)\Theta(J-\Delta^\mathrm{(max)})\,.
\end{equation}
\label{eq:fan}
\end{subequations}
Here $P(h)=\Theta(W-|h|)/2W$
is the uniform disorder distribution and, from Eqs.~\eqref{eq:deltamin} and \eqref{eq:deltamax},  $\Delta^\mathrm{(min)}$ and $\Delta^\mathrm{(max)}$ can be expressed as
\begin{eqnarray}
\Delta^\mathrm{(min)} &=& \Theta(\vert h\vert-J_z)\vert 2h -4\mathrm{sgn}(h)J_z\vert+\Theta(J_z-\vert h\vert)\vert 2h\vert,\nonumber\\
\Delta^\mathrm{(max)}&=&\vert 2h +4\mathrm{sgn}(h)J_z\vert.
\label{eq:deltaminmax}
\end{eqnarray}
This yields
\begin{eqnarray}
f_n &=& \begin{cases}
0 ~~~~~~~~~~~~~~~~~~~~~~~~~~~:~ W<\frac{J}{2}+2J_z\\
\frac{1}{W}\left(W-\frac{J}{2}-2J_z\right)~~~~:~ W\ge\frac{J}{2}+2J_z
\end{cases}\label{eq:fnex}\\
f_a &=& \begin{cases}
1 ~~~~~~~~~~~~~~~~~~~~~~~~~~~:~W\le\frac{J}{2}-2J_z\\
\frac{1}{W}\left(\frac{J}{2}-2J_z\right)~~~~~~~~~~~:~W>\frac{J}{2}-2J_z
\end{cases}\label{eq:faex}
\end{eqnarray}
Upper and lower bounds on the critical disorder are therefore
\begin{equation}
W_c^\pm = \frac{J}{2}\pm2J_z.
\label{eq:wcbounds}
\end{equation}
\begin{figure}[t]
\includegraphics[width=\columnwidth]{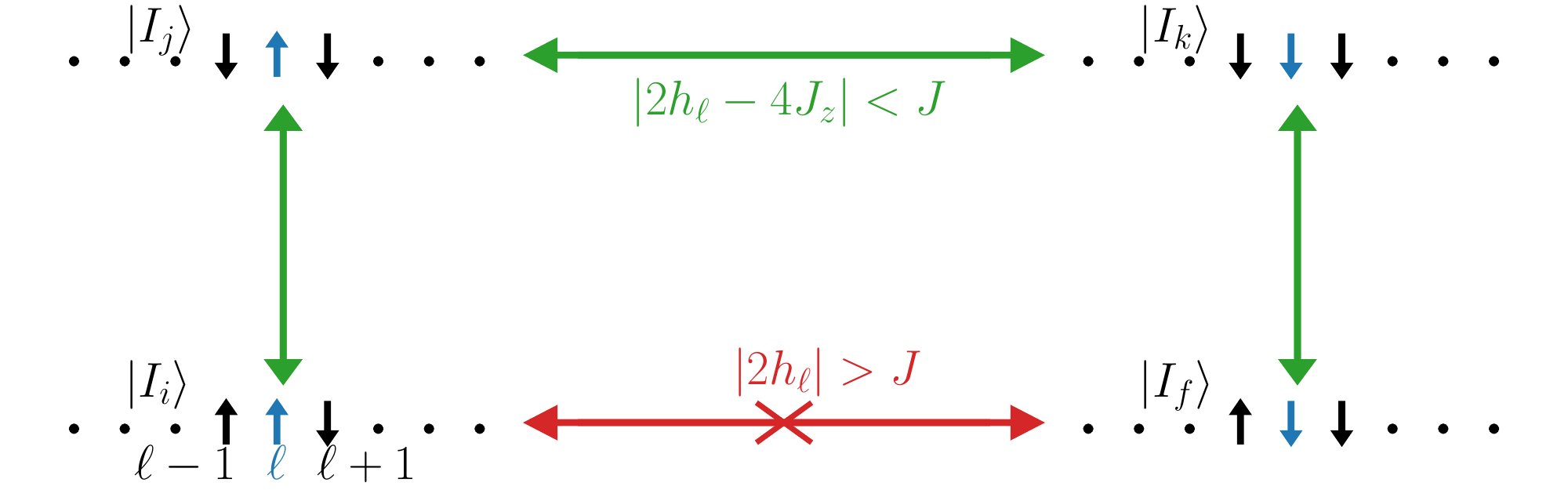}
\caption{Illustration showing that all nodes may belong to a single cluster, even though not all edges satisfy the percolation criterion. The 
spin at site $\ell$ (marked in blue) cannot flip under anti-parallel orientations of its neighbours. This is indicated by the 
red arrows marked with a cross. The nodes $\ket{I_i}$ and $\ket{I_f}$ are therefore not directly connected. These nodes are however connected 
(and hence are part of the same cluster) via a different path, as depicted by the green arrows. This is due to the 
interactions enabling the spin-flip at $\ell$ when both neighbours are down.}
\label{fig:path}
\end{figure}

The behaviour of the three fractions, $f_a$, $f_n$, and $f_s$ is shown in Fig.~\ref{fig:fraction}.
For $W<W_c^-$, $f_a =1$ and the system is trivially percolating.
For $W>W_c^+$, $f_n>0$ and  the system is trivially localised.
This shows that two phases exist in the model, with a disorder-driven transition at a critical 
disorder strength $W_c$ which lies in the range $W_c^-<W_c<W_c^+$.
We add in passing that a bounded disorder distribution is 
required, since for an unbounded distribution $f_n>0$ for any $W>0$,  whence even an infinitesimal disorder would drive the system into the localised phase.

Insight into the nature of this transition comes from noting 
that in the percolating phase close to the critical point at $W_c$,  one has $f_a<1$ and $f_s>0$ (see Fig.~\ref{fig:fraction}).
Hence, some edges on the hypercube are inactive, but there is a macroscopic cluster.
This reflects the fact that there are multiple paths on the hypercube between any two nodes.
As an example consider the case shown in Fig.~\ref{fig:path}. Here, the spin at site $\ell$ cannot be flipped if its neighbours are anti-parallel, and so 
there is no active edge between nodes $\ket{I_i}$ and $\ket{I_f}$. 
On the other hand, this spin can be flipped if its neighbours are both down. The nodes may hence be connected indirectly, and so both belong to the same cluster.
The existence of the indirect path requires non-zero interaction strength $J_z$. 

\begin{figure}
\includegraphics[width=\columnwidth]{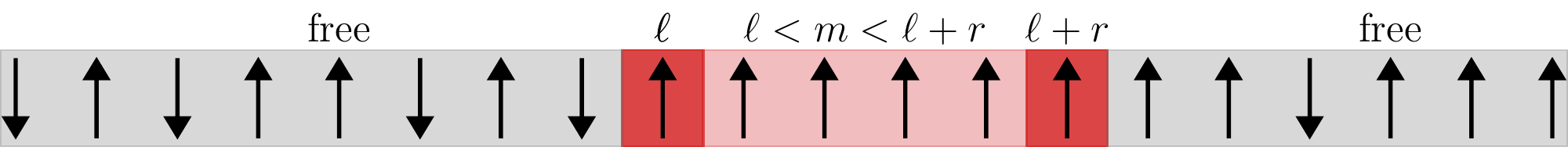}
\caption{Illustration of freezing a local segment of spins in the TFI model. The fields at sites $\ell$ and $\ell+r$ satisfy $h_\ell,h_{\ell+r}>J/2$ whereas the fields at intermediate sites $m$ satisfy the weaker condition $h_m>(J/2)-2J_z$. This results in the entire segment $[\ell,\ell+r]$ being frozen irrespective of the spin configuration of the rest of the system.}
\label{fig:spinfreezingtfi}
\end{figure}

Further insight comes from considering the localised phase close to the critical point at $W_c$. Here $f_n=0$, which indicates that the transition is driven not by the individual freezing of isolated spins but rather by the collective freezing of multiple spins.
As an example, consider a segment of the spin chain of length $r+1$, from site $\ell$ to site $\ell + r$, and ask what conditions must be satisfied by the fields $h_m$ at sites $\ell \leq m \leq \ell + r$ for spins in this segment to be frozen up, regardless of the configurations of other spins in the chain. We require $h_\ell > J/2$ and $h_{\ell +r} > J/2$ at the ends of the segment, and $h_{m} +2J_z > J/2$ in the interior ($\ell < m < \ell + r$).
Thus we have
situation in which two \emph{flag} spins, at sites $\ell$ and $\ell + r$,
are frozen up, irrespective of the configurations of the spins outside the segment (at $\ell-1$ and $\ell + r+1$), as long as the
intermediate spins are also up.
See Fig.~\ref{fig:spinfreezingtfi} for an illustration.

Such a configuration can appear only if $W>J/2$, since $h_\ell\in[-W,W]$. Moreover, the
probability for such a configuration to be present is proportional to
$(W-J/2)^2 \phi^{(r-1)}$ where
$\phi =  \int dh P(h)\Theta(h)\Theta(h-\frac{J}{2}+2J_z)<1$.
This suggests that
the critical disorder strength is $W_c=J/2$ and implies that the finite-size scaling exponent $\nu$ takes the value $2$, since the density of such segments vanishes as $(W-J/2)^2$. It also 
indicates that the probability of a segment of length $r$ being frozen
falls exponentially with increasing $r$.

The Fock-space dimension of a segment of total length $r+1$ is
$2^{r+1}$. If one of the configurations of this segment (the one with all spins up) is frozen, Fock space
automatically splits up into two clusters: one of size (normalised by the
Fock-space dimension) $S/N_\mathcal{H}=2^{-(r+1)}$ and the
other of size $S/N_\mathcal{H}=1-2^{-(r+1)}$. Due to the exponential
suppression factor $\phi^{r-1}$, the most probable instance is
$r=1$. In this case the cluster splits up into
one containing a quarter of the Fock-space sites, and another
containing the remaining three quarters. 
For $W>J/2$, since there is a finite density of such frozen segments, the typical cluster size is a vanishing fraction of the Fock-space dimension in the thermodynamic limit.

\begin{figure}
\includegraphics[width=\columnwidth]{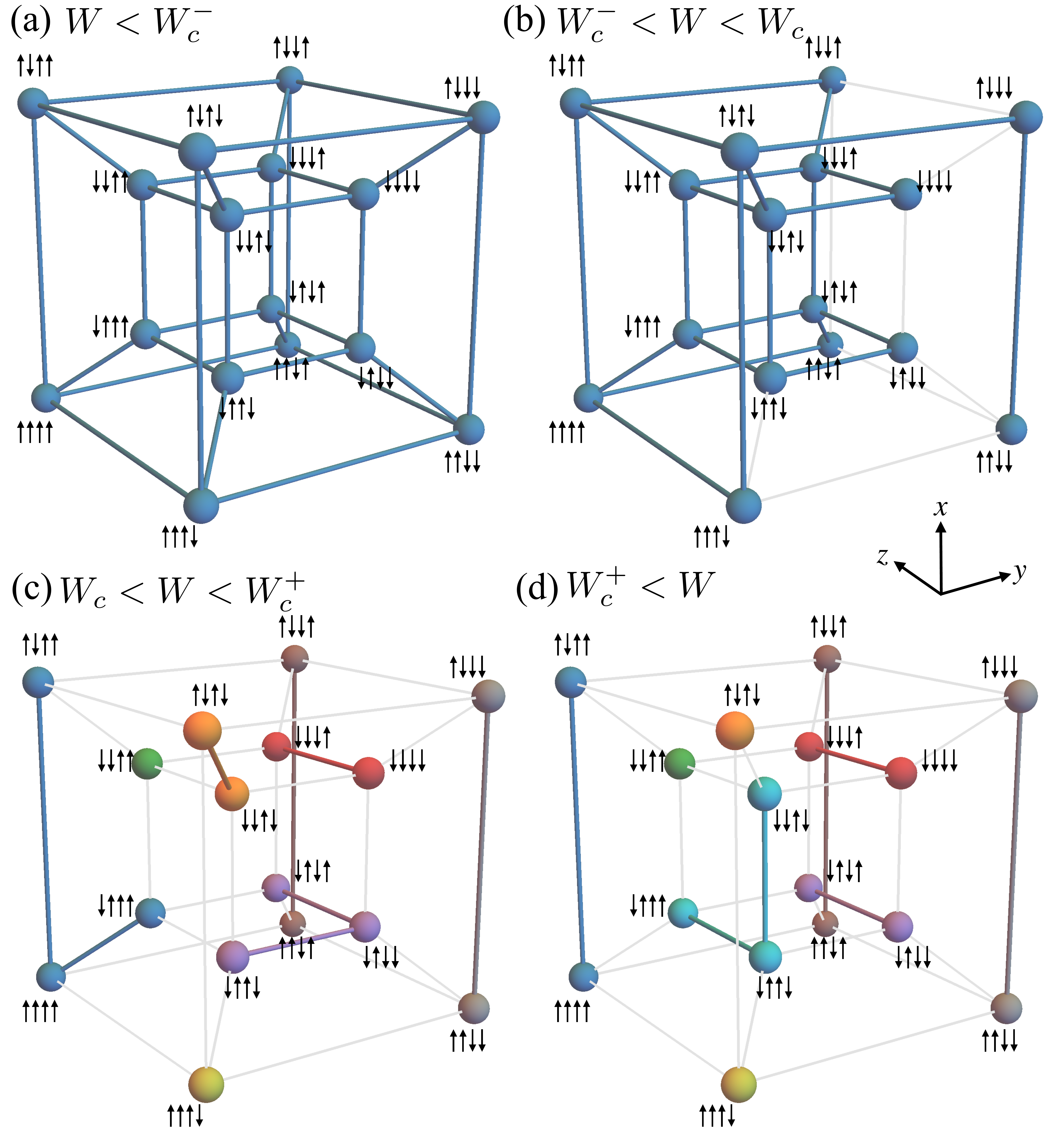}
\caption{Illustration of behaviour in the four regimes of disorder strength, using the Fock-space hypercube for 4 spins. 
All bonds along a particular direction correspond to 
flipping a particular spin.  
Bonds connecting the inner cube to the outer correspond to the first spin, while bonds along $x$, $y$, and $z$ directions correspond to the second, third, and fourth spin respectively.
(a) For $W<W_c^-$, all edges satisfy the percolation criterion. (b) For $W_c^-<W<W_c$, not all edges satisfy the percolation criterion (some blue edges are missing), but all nodes belong to a single cluster. (c) For $W_c<W<W_c^+$ the system is in a localised phase with many clusters, as indicated by the different colours of the nodes. (d) For $W>W_c^+$ some 
spins completely freeze (the first and third in this example), and no bonds along the corresponding directions are active.}
\label{fig:hslattice}
\end{figure}

The picture that we arrive at in this way for behaviour as a function of disorder strength $W$ is exemplified in  Fig.~\ref{fig:hslattice} for a four-site system.

\subsection{XXZ spin chain \label{sec:physicalxxz}}

In the case of the XXZ spin chain [Eq.~\eqref{eq:hamxxz}], hopping on a Fock-space edge corresponds to exchange in real space of oppositely oriented spins at a domain wall. 
Analogously to Eq.~\eqref{eq:deltaap} for the TFI spin chain, the energy cost $\Delta$ for such a hopping depends on the fields on the sites either side of the domain wall (say $\ell$ and $\ell+1$), as well as the spin orientations of their
nearest neighbour sites $\ell-1$ and $\ell+2$. 
The energy costs are 
\begin{align}
\vert\cdot\uparrow\uparrow\downarrow\uparrow\cdot\rangle\leftrightarrow\vert\cdot\uparrow\downarrow\uparrow\uparrow\cdot\rangle\Rightarrow& \Delta_\ell^{(p\uparrow)} = 2\vert h_\ell^\pd-h_{\ell+1}^\pd\vert, \nonumber\\
\vert\cdot\downarrow\uparrow\downarrow\downarrow\cdot\rangle\leftrightarrow\vert\cdot\downarrow\downarrow\uparrow\downarrow\cdot\rangle\Rightarrow& \Delta_\ell^{(p\downarrow)} = 2\vert h_\ell^\pd-h_{\ell+1}^\pd\vert, \nonumber\\
\vert\cdot\uparrow\uparrow\downarrow\downarrow\cdot\rangle\leftrightarrow\vert\cdot\uparrow\downarrow\uparrow\downarrow\cdot\rangle\Rightarrow& \Delta_\ell^{(a_{\uparrow\downarrow})} = 2\vert h_\ell^\pd-h_{\ell+1}^\pd+2J_z^\pd\vert, \nonumber\\
\vert\cdot\downarrow\uparrow\downarrow\uparrow\cdot\rangle\leftrightarrow\vert\cdot\downarrow\downarrow\uparrow\uparrow\cdot\rangle\Rightarrow& \Delta_\ell^{(a_{\downarrow\uparrow})} = 2\vert h_\ell^\pd-h_{\ell+1}^\pd-2J_z^\pd\vert.
\label{eq:deltaapxxz}
\end{align}

These spin exchanges are respectively never or always possible (considering all possible spin configurations on sites $\ell-1$ and $\ell +2$) under the conditions
\begin{eqnarray}
\Delta_\ell^\mathrm{(min)}&=&\min\{\Delta_\ell^{(p\uparrow)},\Delta_\ell^{(p\downarrow)},\Delta_\ell^{(a_{\uparrow\downarrow})},\Delta_\ell^{(a_{\downarrow\uparrow})}\} > J\label{eq:deltaminxxz}\\
\Delta_\ell^\mathrm{(max)}&=&\max\{\Delta_\ell^{(p\uparrow)},\Delta_\ell^{(p\downarrow)},\Delta_\ell^{(a_{\uparrow\downarrow})},\Delta_\ell^{(a_{\downarrow\uparrow})}\} < J\label{eq:deltamaxxxz}
\end{eqnarray}
The probabilities $f_n$ and $f_a$ that a domain never or always flips are
\begin{eqnarray}
f_n &=& \int dh_\ell P(h_\ell)\int dh_{\ell+1}P(h_{\ell+1})\Theta(\Delta_\ell^\mathrm{(min)}-J)\label{eq:fnxxz}\\
{\rm and} &&\nonumber\\
f_a &=& \int dh_\ell P(h_\ell)\int dh_{\ell+1}P(h_{\ell+1})\Theta(J-\Delta_\ell^\mathrm{(max)})\,.\label{eq:faxxz}
\end{eqnarray}
Using Eqs.~\eqref{eq:deltaapxxz}-\eqref{eq:faxxz}, $f_n$ and $f_a$ can be evaluated as

\begin{eqnarray}
f_n &=& \begin{cases}
0 ~~~~~~~~~~~~~~~~~~~~~~~~~~~~~~~~~~:~ W<\frac{J}{4}+J_z\\
\frac{1}{W^2}\left(W-\frac{J}{4}-J_z\right)^2~~~~~~~~~~:~ W\ge\frac{J}{4}+J_z
\end{cases} \label{eq:fnexxxz}\\
{\rm and} &&\nonumber \\
f_a &=& \begin{cases}
1 ~~~~~~~~~~~~~~~~~~~~~~~~~~~~~~~~~~:~W\le\frac{J}{4}-J_z\\
1-\frac{1}{W^2}\left(W-\frac{J}{4}+J_z\right)^2~~~~~:~W>\frac{J}{4}-J_z\,.
\end{cases} \label{eq:faexxxz}
\end{eqnarray}

Hence for $W>J_z+J/4$ we find $f_n>0$. Spins at a finite fraction of domain walls are then unable to flip for any configuration of neighbours. The resulting finite density of frozen spins implies a typical cluster size that is a vanishing fraction of $\nh$. Thus $W_c^+=J_z+J/4$ is an upper bound on the critical disorder strength. 
Similarly, for $W<-J_z+J/4$ we find $f_a=1$. In this case spins at all domain walls can flip for any configuration of neighbours. Thus 
the system is in a trivial percolating phase. Hence $W_c^-=-J_z+J/4$ is a lower bound on the critical disorder strength.

\begin{figure}
\includegraphics[width=\columnwidth]{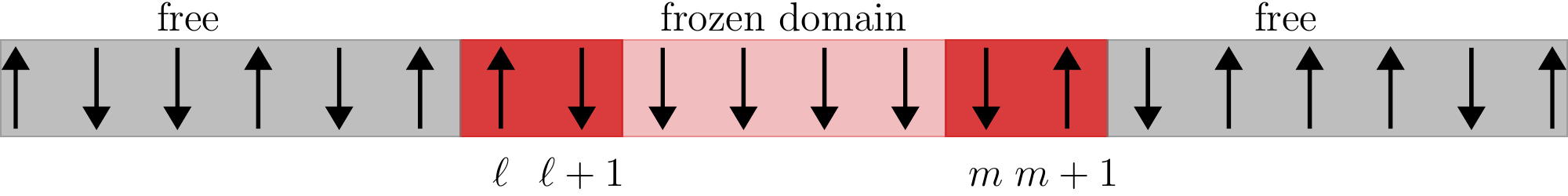}
\caption{A schematic example of how a domain of spins can freeze in the XXZ model. In this example of a down-spin domain, the fields at the left end satisfy $|h_\ell - h_{\ell+1} |>J/2$ and $|h_\ell - h_{\ell+1}+2J_z |>J/2$. Similarly, those at the right end satisfy $|h_{m +1} - h_{m} |>J/2$ and $|h_{m +1} - h_{m}+2J_z |>J/2$. The spins within the domain are frozen irrespective of the values of $h_k$ for $\ell+1 < k < m$, and irrespective of the spin configuration on sites 
outside the domain ($k<\ell$ or $k>m+1$).}
\label{fig:spinfreezingxxz}
\end{figure}

As for the TFI model, a simple picture of the transition can be constructed for the XXZ model by analysing the conditions under which spins freeze.
To describe this picture, 
consider for definiteness the second and third lines of Eq.~(\ref{eq:deltaapxxz}) and suppose 
that $\vert h_\ell^\pd-h_{\ell+1}^\pd\vert > J/2$ and $\vert h_\ell^\pd-h_{\ell+1}^\pd+2J_z^\pd\vert>J/2$. Then a domain 
of down spins that has its left-hand end at site $\ell +1$ is stable for both possible orientations of the spin at site 
$\ell-1$. Assume (to be specific) that $h_\ell^\pd-h_{\ell+1}^\pd >0$. Then the probability of satisfying the first 
condition is $\int dhP(h)\int dh^\prime P(h^\prime)~\Theta(h-h^\prime-J/2)\sim (W-J/4)^2$ for small
$W-J/4>0$, and when it is satisfied, so too
is the second condition. There will be a finite density of such domain walls in the thermodynamic limit, and
analogous circumstances will provide a stable right-hand end to the domain of down spins (while 
domains of up spins can be formed in an equivalent way). See Fig.~\ref{fig:spinfreezingxxz} for an illustration.

There is important difference between these domains in the XXZ model and the frozen sequences between flag spins in the TFI model, because spin flips in the XXZ model can only ever take place at domain walls, while in the TFI model they may occur at any site if the relevant $\Delta$ is not too large. The domains in the XXZ model are hence stable regardless of the fields acting on sites in the domain interior, whereas the sequence between flag spins in the TFI model is frozen only if the fields lie outside a threshold. Because of this, long domains are not penalised in the XXZ model. Instead, their typical length is set by the density of stable domain ends, implying a critical disorder  strength $W_c = J/4$, and again an exponent 
of $\nu=2$.

In summary, although the value of $\nu$ obtained for both models is the same, the detailed picture of the localised phase is quite different in the two cases. 
In the TFI model, short segments of spins freeze while all others remain flippable. 
By contrast, in the XXZ model
rather large real-space patches freeze.

\section{Numerical results \label{sec:numerical}}

In this section, we present numerical results to test and corroborate the preceding arguments for the existence of two phases and the nature of the transition between them.
In Ref.~\cite{roy2018exact}, numerical results for average/typical cluster sizes were shown for the TFI model.
To illustrate the generality of the approach, 
we describe here results from numerical calculations on the XXZ model Eq.\ \eqref{eq:hamxxz}, which has the practical numerical advantage of having a smaller Fock-space $\nh={{N}\choose{N/2}}$ compared to the TFI spin chain ($\nh=2^N$) for the same $N$, owing to conserved total magnetisation.

The distribution of $\mathcal{E}$ over the Fock space is Gaussian with a standard deviation which scales 
as $\sqrt{N}$~\cite{welsh2018simple}.
We focus on the cluster that contains the node $|I_0\rangle$ whose energy is closest to the mean of 
the $\mathcal{E}$, since that is where the distribution 
peaks.
In other words, for a given disorder realisation, $\ket{I_0}$ is identified as the node which satisfies 
$\mathcal{E}_{I_0} = \mathrm{min}_I\{\vert \mathcal{E}_I-\sum_I\mathcal{E}_I/N_\mathcal{H}\vert\}$.
Starting from the node $\ket{I_0}$, we grow the cluster containing it (henceforth denoted as $\mathcal{C}$) 
following the percolation rules described in Sec.~\ref{sec:percolationmodel}.
Over many disorder realisations, we obtain an ensemble of clusters of sizes $N_\mathcal{C}$,  
and examine their statistical properties.

Note that 
for all edges on the Fock-space to be active in the limit $W=0$
requires $J>4J_z$, and we make this restriction in all numerical calculations.
In particular we take $J_z=1$ and $J=4.1$. 
We emphasise that this is not a fine-tuned parameter choice, but merely ensures 
a finite, albeit narrow, regime of disorder $W<W_c^-$ supporting the trivial percolating phase.
We recall from Sec.~\ref{sec:physicalxxz} the 
argument that $W_c = J/4 = 1.025$ for the disordered XXZ chain. This is consistent with various numerical 
diagnostics given in Secs.~\ref{sec:meantypical}-\ref{sec:szcluster}.

\subsection{Distribution of cluster sizes \label{sec:distributions}}
\begin{figure*}
\includegraphics[width=\linewidth]{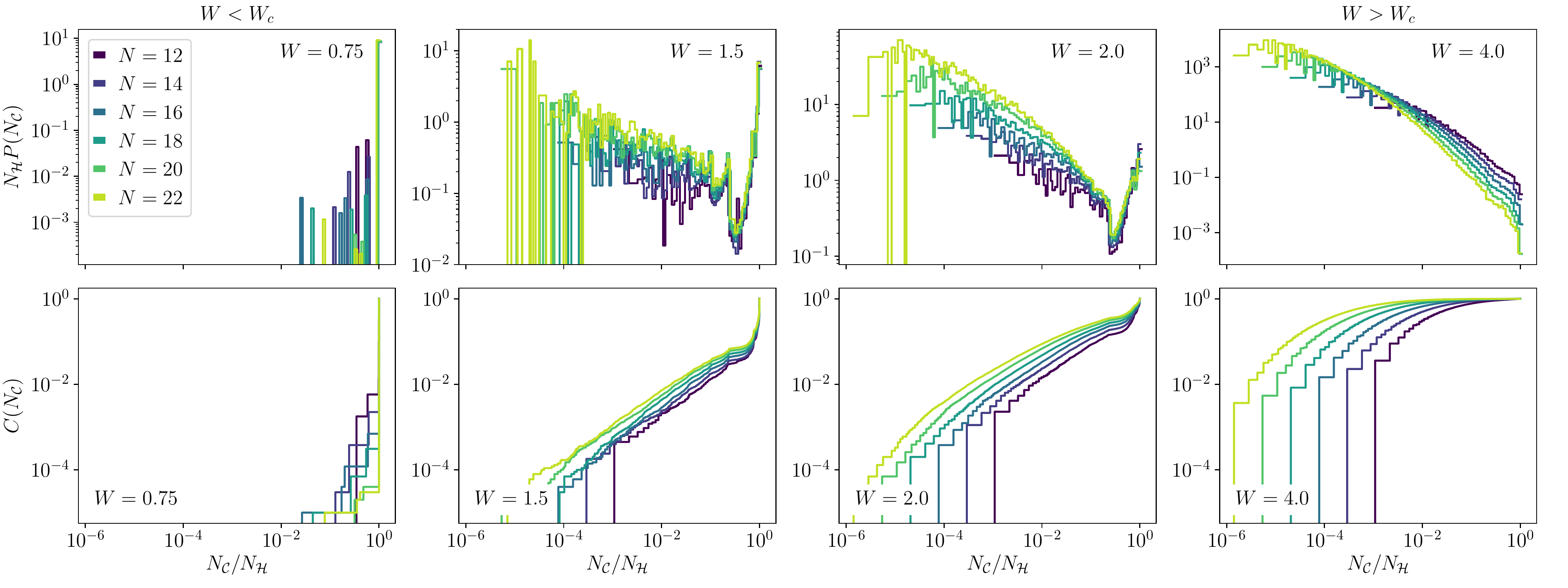}
\caption{Distributions for the disordered XXZ model ($W_c =1.025$ for the parameters chosen, as is confirmed  via finite-size scaling analyses). Top row: probability distributions of cluster sizes, $\tilde{P}(\nc/\nh)=\nh P(\nc)$, for different system sizes $N$ 
and disorder $W$ (as indicated in panels).
Bottom row: corresponding cumulative distributions $C(\nc)$.
For $W<W_c$ almost the entire weight of the distributions is on $\nh/\nc=1$, whereas for 
strong disorder the weight shifts towards to $\nh/\nc\to 0$ as $N$ increases. For $W>W_c$, but not too deep inside the localised phase, the distribution for finite $N$ has a bimodal nature with its weight depleting from $\nc=\nh$ to clusters of sizes $\nc\sim\nh^\alpha$ with $\alpha <1$. However the peak at $\nc=\nh$ loses weight with increasing $N$, suggesting that it 
vanishes in the thermodynamic limit.
See text for discussion. For the numerics, $J_z=1$ and $J=4.1$, with statistics obtained from $10^5$ realisations. }
\label{fig:sizedist}
\end{figure*}

We start by discussing the distribution $P(N_\mathcal{C})$ of cluster sizes, to present a broad 
qualitative picture of the two phases and the transition
between them.
A related quantity is $p(s)$, the probability over all clusters of there being a cluster of size $s$. 
In the percolating phase, $p(s)$ may not reveal in a simple way the presence of the percolating cluster with 
$s\sim N_\mathcal{H}$, because
there can exist only $\mathcal{O}(1)$ such clusters, but in addition there may exist $\mathcal{O}(N_\mathcal{H}^{1-\alpha})$ clusters with $s\sim N_\mathcal{H}^\alpha$.
Since $N_\mathcal{H}^{1-\alpha} \gg 1$ in the thermodynamic limit, the distribution will be dominated by the 
smaller clusters.
In percolation theory, this issue is taken care of by looking at the probability that a chosen site is in a cluster of size $s$~\cite{stauffer2014introduction}.
This probability, denoted as $P(s)$, is given by 
\begin{equation}
P(s) = \frac{sp(s)}{\int ds~s p(s)}.
\label{eq:P(s)}
\end{equation}
Since we study the particular cluster $\mathcal{C}$ containing the node 
$\ket{I_0}$, the distribution obtained over disorder realisations is automatically $P(N_\mathcal{C})$.
In addition, we also consider the cumulative distribution $C(N_\mathcal{C})$, defined by
\begin{equation}
C(s) = \int_1^s ds^\prime P(s^\prime),
\label{eq:C(s)}
\end{equation}
since certain aspects of the statistics are more clearly seen with it.
To compare different system sizes, we also
find it useful to study the data as a function of $N_\mathcal{C}/N_\mathcal{H}$.
Hence, we define the corresponding normalised distribution as $\tilde{P}(\nc/\nh)=\nh P(\nc)$.

Representative results for $\tilde{P}(N_\mathcal{C}/N_\mathcal{H})$ and $C(N_\mathcal{C})$ are shown 
in Fig.~\ref{fig:sizedist}.
Two general aspects of the distributions are important.
First, how they behave as a function of system size for a fixed disorder. 
Second, how the behaviour with system size 
evolves as the disorder is tuned across the transition.
Disentangling the two aspects is quite subtle, and in the following we elaborate on how the natures of the percolating and localised phases are 
embodied in the distributions shown in Fig.~\ref{fig:sizedist}.

For disorder below the critical $W_{c}$,
it is clear that almost all the weight of the distribution is at $N_\mathcal{C}/N_\mathcal{H} = 1$ and
that this weight tends towards unity with increasing system size $N$. 
As an example, consider the panels corresponding to $W=0.75$ in Fig.~\ref{fig:sizedist}.
With increasing $N$, the weight of the distributions at $\nc/\nh<1$ decreases, as indicated by 
the progressive disappearance of the secondary peaks in $\tilde{P}(\nc/\nh)$.
This is more clearly seen in the cumulative distributions, where the curve corresponding to a larger $N$ is always below 
one corresponding to a smaller $N,$ indicating that the cumulative weight of the distribution at all $\nc<\nh$ decreases 
with increasing system size.
Hence from the trends it can be 
inferred that in the thermodynamic limit
$\tilde{P}(\nc/\nh)\to \delta(\nc/\nh-1)$, which is a defining feature of the 
percolating phase.

In the opposite limit of 
large disorder $W>W_c^+$, as discussed in Sec.\ \ref{sec:physical}, a finite fraction of the spins are anyway 
guaranteed completely frozen and hence there is no possibility of a percolating cluster. 
This is also corroborated by the data (see the panels corresponding to $W=4$).
The weight of $\tilde{P}(\nc/\nh)$ monotonically decreases with increasing $N$ for cluster sizes which are finite fractions of the Fock-space, and consequently increases for vanishing fractions since the distributions are 
normalised.
Further, the cumulative distribution also shows the same behaviour: $C(\nc)$ saturates to unity at smaller values of 
$\nc/\nh$ as $N$ increases, thus showing that the 
full weight of the distributions is on clusters whose sizes are a 
vanishing fraction of the Fock-space dimension.

We now discuss the nature of the distributions at intermediate disorder, specifically in the vicinity of the critical point.
On increasing the disorder above $W_{c}$,
for a finite system size, the peak at $\nc/\nh\approx 1$ in $\tilde{P}(\nc/\nh)$ begins to lose weight and the distribution gains weight at smaller values of $\nc/\nh$.
Some further important observations 
can be made from the data.

First, for disorder $W>W_c$, but not too 
large, there is a perceptible peak in $\tilde{P}(\nc/\nh)$ at $\nc/\nh=1$ and a corresponding jump in $C(\nc)$. 
However, the weight of the distribution in this peak decreases with both increasing $N$ and $W$ (see the panels corresponding to $W=1.5$ and $W=2$ in Fig.~\ref{fig:sizedist}). 
While the behaviour with increasing $W$ is quite clear in the data for $\tilde{P}(\nc/\nh)$, it is less so with increasing $N$ 
due to the relatively small system sizes accessible numerically.
It is however clearer in the data for $C(\nc)$, where the curves for larger $N$ are always above those for smaller $N$.
This indicates that with increasing $N$, the distributions have more weight on smaller values of $\nc/\nh$.
And since the distributions $\tilde{P}(\nc/\nh)$ are normalised, 
this means that the peak at $\nc/\nh=1$ loses weight with increasing $N$, and as such 
vanishes in the thermodynamic limit.
This behaviour can be traced to a continuously varying exponent, $\alpha$, which describes the scaling of mean/typical cluster sizes with the Fock-space dimension, and which we introduce and consider
in the next sub-section (\ref{sec:meantypical}).

Second, the aforementioned peak only loses weight with increasing $N$ or $W$, but does not shift to a finite value of $\nc/\nh$.
One can 
infer from this observation that it is very rare 
(vanishingly so in the thermodynamic limit) that the Fock space has multiple clusters of sizes which are a finite fraction of $\nh$. In other words, if there exists a cluster whose size is a finite fraction of $\nh$, it 
covers the entire Fock-space in the thermodynamic limit.
This indicates that the percolating cluster close to the phase transition
does not fragment into multiple smaller clusters of sizes which are finite fractions of $\nh$,
but rather into 
clusters all of whose sizes are vanishing fractions of the Fock-space dimension. This is consistent with the physical picture presented in Sec.~\ref{sec:physicalxxz}.

The distributions $\tilde{P}(\nc/\nh)$ and $C(\nh/\nh)$ thus 
show qualitatively different behaviour for weak and strong disorder, including opposite trends with increasing system size, and provide 
clear initial evidence that there 
exists a phase transition in the model. 
The data also corroborate 
the physical picture presented for the transition in Sec.~\ref{sec:physical}.

\subsection{Typical and mean cluster sizes \label{sec:meantypical}}

To extract quantitative information about the critical disorder and the finite-size scaling exponent, we now consider mean cluster sizes.
 Specifically, we compute both the arithmetic mean 
 and the geometric mean, 
 given respectively by
\begin{eqnarray}
\mathcal{S}_\mathrm{avg} &=& \int_1^{\nh} ds~s P(s)\label{eq:savg}
\end{eqnarray}
and
\begin{eqnarray}
\mathcal{S}_\mathrm{typ} &=& \exp\left[\int_1^{\nh} ds \log(s) P(s)\right].\label{eq:styp}
\end{eqnarray}
From the nature of the distributions shown in Fig.~\ref{fig:sizedist}, $\mathcal{S}_\mathrm{typ}$ or
$\mathcal{S}_\mathrm{avg}$
appear to be natural diagnostics for the phase transition, as indeed is shown below. One can also draw a parallel 
between $\mathcal{S}_\mathrm{typ}$
and participation entropies of the many-body eigenstates of the quantum system, which serve as useful diagnostics for 
the many-body localisation transition~\cite{deluca2013ergodicity,luitz2015many}.
For an eigenstate $\ket{\psi}$, the first participation entropy in the Fock space is defined as 
$S_1(\ket{\psi}) = -\sum_I \vert \braket{\psi|I}\vert^2\log\vert\braket{\psi|I}\vert^2$.
The disorder-averaged $S_1$ has a general form~\cite{deluca2013ergodicity,luitz2015many}
\begin{equation}
S_1 = \alpha\log\nh + \beta\log(\log\nh),
\label{eq:participationentropy}
\end{equation}
with $\alpha =1$ in the delocalised phase, indicating that eigenstates have on average support on the entire 
Fock-space; and  with $\alpha <1$ in the many-body localised phase, indicating that eigenstate support 
exists on a vanishing fraction $\sim N_{{\cal{H}}}^{\alpha -1}$ of Fock-space.
Note that the probability density $p_I^\pd=\vert\braket{\psi|I}\vert^2$ can be interpreted as the probability or the weight of the eigenstate on the basis state $\ket{I}$, normalised over those basis states on which the eigenstate has 
support.
In our classical problem, any node is either a member of the cluster $\mathcal{C}$ or not, and from the set of nodes $I\in\mathcal{C}$, the probability that a randomly chosen node is a particular node $I$ is (trivially) $p^\pd_I=1/\nc$. 
Hence $p_I^\pd$ is the weight of the cluster on the node $I$ normalised over the nodes contained in the cluster, 
directly analogous to $\vert\braket{\psi|I}\vert^2$ in the quantum problem.
Formally, this can be expressed by defining for a given disorder realisation the scaled indicator function
\begin{equation}
p_I^\pd = \begin{cases}
		1/\nc~~~~:~I\in\mathcal{C}\\
		0 ~~~~~~~~~:~I \notin \mathcal{C},
	\end{cases}
\label{eq:AI}
\end{equation}
so that $\sum_Ip_I^\pd=1$ 
The equivalent of the first participation entropy is then 
\begin{equation}
S_1^\mathcal{C} = -\overline{\sum_I p_I\log p_I} ~=~ \overline{\log\nc}
\label{eq:participationentropyanalogue}
\end{equation}
where the overline denotes disorder averaging.
Comparing Eq.~\eqref{eq:participationentropyanalogue} with Eq.~\eqref{eq:styp} 
shows directly that the analogue of the first participation entropy is simply the logarithm of the typical cluster size.

\begin{figure}
\includegraphics[width=\linewidth]{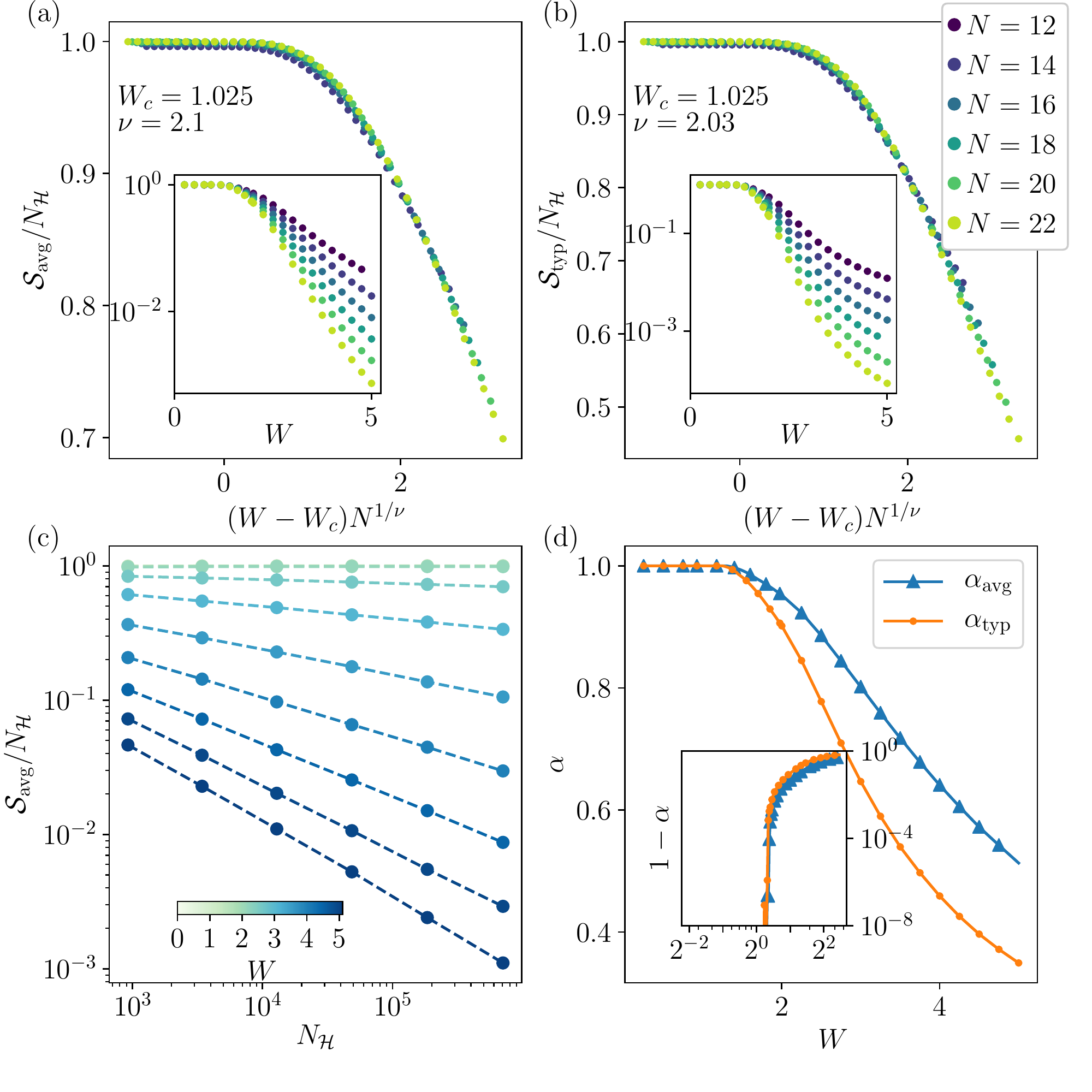}
\caption{Results for the disordered XXZ chain. (a)~$\mathcal{S}_\mathrm{avg}/\nh$ and (b)~$\mathcal{S}_\mathrm{typ}/\nh$ 
versus disorder $W$, for system sizes $N=12-22$. Insets show raw data whereas main panels show data collapsed onto 
a common function of $(W-W_c)N^{1/\nu}$, with $W_c$ constrained to $J/4$ and the extracted value of $\nu$ given in the panels. (c)~$\mathcal{S}_\mathrm{avg}/\nh$ as a function of $\nh$ for various values of $W$ as indicated by the colour scale.
Specifically, the set of values are $\{0.25,0.75,1.75,2.25,2.75,3.25,3.75,4.25,4.75\}$.
$\mathcal{S}_\mathrm{avg}/\nh =1$ arises for $W<W_c$ (two such $W$ values are shown: 0.25 and 0.75; they are indistinguishable in this figure and 
both give $\alpha =1$).
Circles are data points and dashed lines are fits to the form Eq.~\eqref{eq:Savgtypfit}, with the extracted values 
of $\alpha$ shown in panel (d) for both $\mathcal{S}_\mathrm{avg}$ and $\mathcal{S}_\mathrm{typ}$. The inset in (d) shows the approach of $\alpha$ to 1 as the critical disorder is
approached from the localised side. $J_z=1$ and $J=4.1$, with averaging over $10^5$ realisations. Statistical errors are calculated from 500 bootstrap resamplings, and are not shown as they are smaller than the data markers.}
\label{fig:meantypical}
\end{figure}

The results for $\mathcal{S}_\mathrm{avg/typ}$ are shown in Fig.~\ref{fig:meantypical}.
It is clear that for disorder $W$ smaller than a critical $W_c$, $\mathcal{S}_\mathrm{avg/typ}/\nh\to1$ (there are extremely small finite-size effects, with $\mathcal{S}_\mathrm{avg/typ}/\nh$ increasing with $N$ but saturating 
to one).
For $W>W_c$ by contrast, $\mathcal{S}_\mathrm{avg/typ}/\nh$ systematically decreases with increasing $N$, as illustrated by the insets to
Fig.~\ref{fig:meantypical}(a) and \ref{fig:meantypical}(b). 

The values of
$W_c$ and the scaling exponent $\nu$ can be obtained by collapsing the data for various system sizes onto a common scaling function of form $g_\mathcal{S}^\pd[(W-W_c)N^{1/\nu}]$.
We perform the finite-size scaling analysis by letting both $W_c$ and $\nu$ be fitting-parameters, as well as constraining one of them to its value suggested by the arguments given in Sec.~\ref{sec:physicalxxz} (namely $W_c=J/4$
and $\nu=2$) and treating the other as a fitting-parameter.
The scaling collapse of the data with the constraint $W_c=J/4 =1.025$ is shown in the main panels of 
Fig.~\ref{fig:meantypical}(a) and \ref{fig:meantypical}(b). This yields $\nu=2.1$ and $2.03$ from the mean and typical cluster sizes respectively, 
which are close both to each other and to the analytically predicted value of 
2~\cite{roy2018exact}.
Allowing both $W_c$ and $\nu$ as fitting-parameters yields $W_c = 1.04$ and $W_c = 0.95$ from the mean and typical cluster sizes respectively, and $\nu=1.9$ from both.
The uncertainty in $W_c$ obtained from the fits can be estimated by constraining $\nu=2$ and extracting the value 
of $W_c$. Such an analysis yields $W_c=1.01$ and $W_c=0.99$ from the average and typical data respectively, and thus uncertainties can be estimated as $\delta W_c=0.03$ and $\delta W_c=0.07$ respectively.

The above goes quite 
some way in showing the universality of $\nu=2$, since this value is obtained by numerical calculations on two different models and analytic 
results for one of them. It is also consistent with the physical arguments given in 
Sec.\ \ref{sec:physical}.

We also emphasise that the exponent $\nu=2$ conforms to the Harris-CCFS bounds~\cite{harris1974effect,chayes1986finite}, which state that for a disorder-driven transition with a length scale that diverges as the critical point is approached from the localised side, the correlation length exponent $\nu$ satisfies the inequality  $\nu\ge 2/d$ where $d$ is the spatial dimension~\cite{chayes1986finite}.
The derivation of such bounds has also been extended 
to many-body localisation transitions~\cite{chandran2015finite}.

$\mathcal{S}_\mathrm{avg/typ}$ can further be analysed by studying their scaling with the Fock-space dimension.
Taking a cue from the scaling of participation entropies, Eq.~\eqref{eq:participationentropy}, we fit $\mathcal{S}_\mathrm{avg/typ}$ to a form 
\begin{equation}
\mathcal{S}_\mathrm{avg/typ} = c_\mathrm{avg/typ}\nh^{\alpha_\mathrm{avg/typ}} .
\label{eq:Savgtypfit}
\end{equation}
The data and fits for a representative range $W$ are shown in Fig.~\ref{fig:meantypical}(c).
The clear linear behaviour of $\mathcal{S}_\mathrm{avg/typ}$ as a function of $\nh$ on logarithmic axes validates the fitting form used in Eq.~\eqref{eq:Savgtypfit}.
The extracted value of $\alpha$ is shown as a function of disorder $W$ in Fig.~\ref{fig:meantypical}(d).
For $W<W_c$, $\alpha=1$ (two such $W$ values are in fact shown in Fig.~\ref{fig:meantypical}(c), both giving $\alpha =1$).
This indicates that in the percolating  
phase, the cluster contains a finite-fraction of the Fock-space nodes.
By contrast, $\alpha<1$ for $W>W_c$, showing
that the cluster is supported over only a vanishing fraction $\mathcal{O}(\nh^{\alpha-1})$ of the Fock-space.
This is similar to the 
behaviour shown by many-body localised eigenstates in the Fock space via their participation entropies~\cite{luitz2015many} and inverse participation ratios~\cite{deluca2013ergodicity}.
Note that 
the form of $\mathcal{S}_\mathrm{avg/typ}$ in Eq.~\eqref{eq:Savgtypfit} also determines the functional form of the scaling function, $g^\pd_\mathcal{S}[(W-W_c)N^{1/\nu}]$, as $g^\pd_\mathcal{S}(x)\sim \exp(-\gamma x^\nu)$ (with $\gamma$ a constant), and further shows that $\alpha$  approaches unity as $(1-\alpha) \propto (W-W_c)^\nu$ when the transition is approached from the localised side.

We also add that the extent of the finite-size effects near the transition for $W>W_{c}$, as discussed in
Sec.\ \ref{sec:distributions}, can be understood by noting that $\alpha$ tends to 1 \emph{continuously} as
$W_c$ is approached from the localised side. 
As a consequence, finite-size effects are naturally large close to the transition:
even for values of $\alpha \approx 0.9$ -- which is well 
within the localised phase, and for which $\nh^{\alpha -1}$ vanishes in the thermodynamic limit --
$\nh^{\alpha -1}\approx 0.5$ remains a finite and sizeable fraction for the largest system sizes accessible to us 
in practice ($\nh \approx 10^{6}$).

Finally, the approach of $\alpha$ to 1 as $W \rightarrow W_c$ from the localised side should also 
consistently indicate the critical point.
To show this, $1-\alpha$ is plotted against $W$ on logarithmic axes in the inset to
Fig.~\ref{fig:meantypical}(d), and
is indeed seen to vanish at $W\simeq 1$, rather close to the 
$W_c$ obtained from the scaling collapse.

\subsection{Fluctuations in cluster sizes \label{sec:fluctuations}}
\begin{figure}
\includegraphics[width=\linewidth]{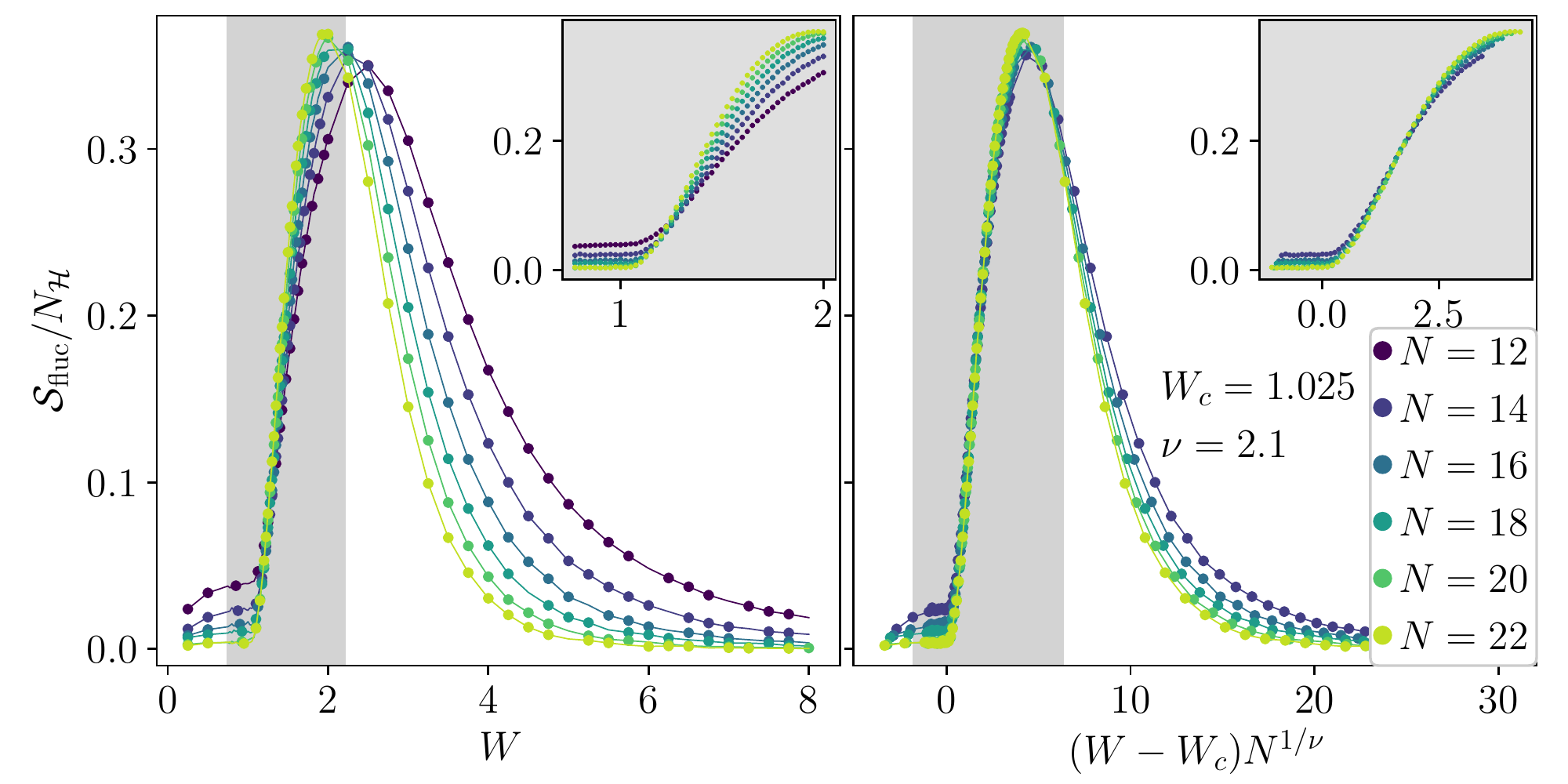}
\caption{Left panel: raw data for $\mathcal{S}_\mathrm{fluc}/\nh$ (Eq.~\eqref{eq:sfluc}) as a function of disorder $W$, for various system sizes $N$. Inset: zoom of the critical region, corresponding to the grey shaded region in the main panel, showing crossing of the data for various $N$. Right panel: the same data scale-collapsed as
a function of $(W-W_{c})N^{1/\nu}$, with $\nu$ the finite-size scaling exponent. 
}
\label{fig:fluctuations}
\end{figure}

The mean/typical cluster sizes, $\mathcal{S}_\mathrm{avg/typ}$, show a volume law in the Fock space
($\propto \nh$) in the percolating phase, and a transition to a sub-volume law $\propto \nh^{\alpha}$ ($\alpha <1$) in 
the localised phase; hence acting as a diagnostic for the phase transition. It is thus natural to look at fluctuations of 
the cluster sizes, defined by
\begin{equation}
\mathcal{S}_\mathrm{fluc} = \left[\int_1^{\nh} ds~s^2P(s) - \left(\int_1^{\nh} ds~sP(s)\right)^2\right]^{1/2}.
\label{eq:sfluc}
\end{equation}
One expects $\mathcal{S}_\mathrm{fluc}$ as a function of $W$ to show a peak at the critical disorder strength.
Physically, this reflects the expectation that in the thermodynamic limit,  fluctuations in 
$N_{\mathcal{C}}/N_{\mathcal{H}}$ are essentially absent in either phase, but show a peak at the critical point.
The latter will naturally be broadened in a finite-sized system, with  the resultant ensemble of cluster sizes expected to contain sizes representative of both phases, and hence to show large fluctuations.
This is also reflected in the fact (see Fig.~\ref{fig:sizedist}) that the distributions $\tilde{P}(\nc/\nh)$
show bimodal character near the transition.

The numerical results shown in Fig.~\ref{fig:fluctuations} indeed corroborate the discussions above.
For sufficiently weak or strong disorder the fluctuations decay systematically with increasing system size, as within a phase fluctuations should not occur in the thermodynamic limit.
Near the critical point by contrast, $\mathcal{S}_\mathrm{fluc}$ shows a peak which becomes sharper with increasing $N$, thus naturally leading to a crossing of the data for various $N$.
A finite-size scaling analysis similar to that for $\mathcal{S}_\mathrm{avg/typ}$ can also be performed, and the data for various system sizes collapsed onto a universal scaling function of $(W-W_c)N^{1/\nu}$.
Such a scaling collapse is shown in Fig.~\ref{fig:fluctuations}(right), 
where we constrain $W_c=J/4$. This yields $\nu=2.1$, 
again close to the expected universal value. Using both $W_c$ and $\nu$ as fitting-parameters yields $W_c=1.07$ and an exponent $\nu=1.8$, 
quite close to those obtained from other diagnostics. 
As in Sec.~\ref{sec:meantypical}, to obtain the uncertainty in $W_c$ we constrain $\nu=2$ and extract $W_c$, 
giving $W_c =1.01$, and thus suggesting 
an uncertainty of $\delta W_c=0.06$.

Finally here, we remark that while there is no direct analogue of entanglement entropy in the classical problem, the peak in $\mathcal{S}_\mathrm{fluc}$ at the critical point is reminiscent of the peak in fluctuations of entanglement entropy at the many-body localisation transition~\cite{luitz2015many,luitz2016long,kjall2014many}; with the entanglement entropy itself showing a volume-law to area-law transition across the critical disorder.

\subsection{Local magnetisation in cluster \label{sec:szcluster}}

\begin{figure}
\includegraphics[width=\linewidth]{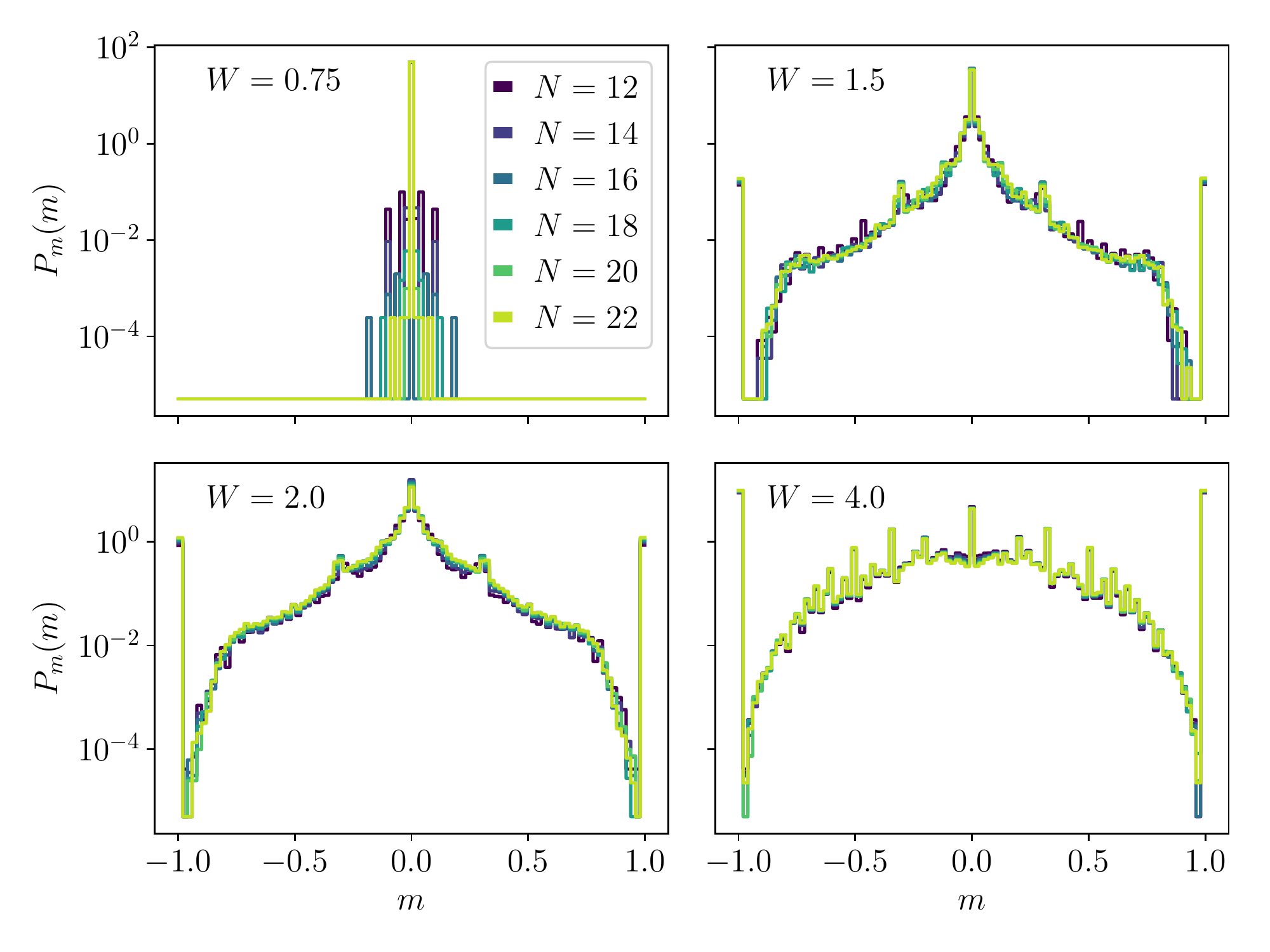}
\caption{Distribution of local magnetisation of the cluster, $P_m(m)$, for four different disorder values, $W$, and different system sizes $N$ (specified in panels). For weak disorder $P_m(m)$ is strongly peaked at $m=0$, and becomes sharper with increasing $N$. On increasing $W$, additional peaks at $m\pm1$ appear, and eventually dominate the distribution at strong disorder. 
}
\label{fig:spindist}
\end{figure}

We now show that physical observables which are local in real space, but measured on the clusters, also carry signatures 
of the transition. 
While the entire percolation problem is set up on the Fock space, being able to make a connection to local observables is 
important, because a fundamental aspect of many-body localised phases is that local observables 
violate the eigenstate thermalisation hypothesis. 
Statistical properties of eigenstate expectation values of local observables have indeed been studied, and shown to distinguish the many-body localised phases from their ergodic counterparts~\cite{baldwin2016manybody,luitz2016long}.

In our classical problem, since the nodes of the Fock-space graph are simply product states with $\sigma^z=\pm1$, the local magnetisation $\sigma^z_\ell$ is a natural choice for the observable.
We define the average local magnetisation of the cluster as 
\begin{equation}
m_\ell^\pd = \sum_I p_I^\pd\bra{I}\sigma^z_\ell\ket{I},
\label{eq:szcluster}
\end{equation}
where $p_I$ is defined in Eq.~\eqref{eq:AI}.
The motivation behind the definition Eq.~\eqref{eq:szcluster} can be understood by considering two limiting cases.
In the limit of very weak disorder where all nodes of the Fock-space graph are in the cluster, 
$m_\ell$ reduces to the trace of the operator $\sigma^z_\ell$ and hence vanishes.
Consequently, the distribution of $m_\ell$ taken over all sites $\ell$ and many disorder 
realisations, $P_m(m)$,
is simply a $\delta$-function at $m=0$. 
In the opposite extreme of very strong disorder, the cluster $\mathcal{C}$ typically has a single node and hence $m_\ell=\pm 1$, resulting in a bimodal distribution with peaks at $m=\pm 1$.
The distributions in the two limits are thus qualitatively different.
With this understanding, we expect that $P_m(m)$ can be written in the thermodynamic limit as a sum of $\delta$-functions and a smooth part, in the form
\begin{equation}
P_m(m)=a_0\delta(m)+a_1\delta(\vert m\vert-1)+P_\mathrm{smooth}(m).
\label{eq:pmdelta}
\end{equation}
In the percolating phase it is expected that $a_0=1$ and $a_1=0$, while in the localised phase $a_0$ and $a_1$ deviate from 1 and 0 respectively.
In the extreme limit $W\to\infty$, one anticipates $a_0=0$ and $a_1=1$.

\begin{figure}
\includegraphics[width=\columnwidth]{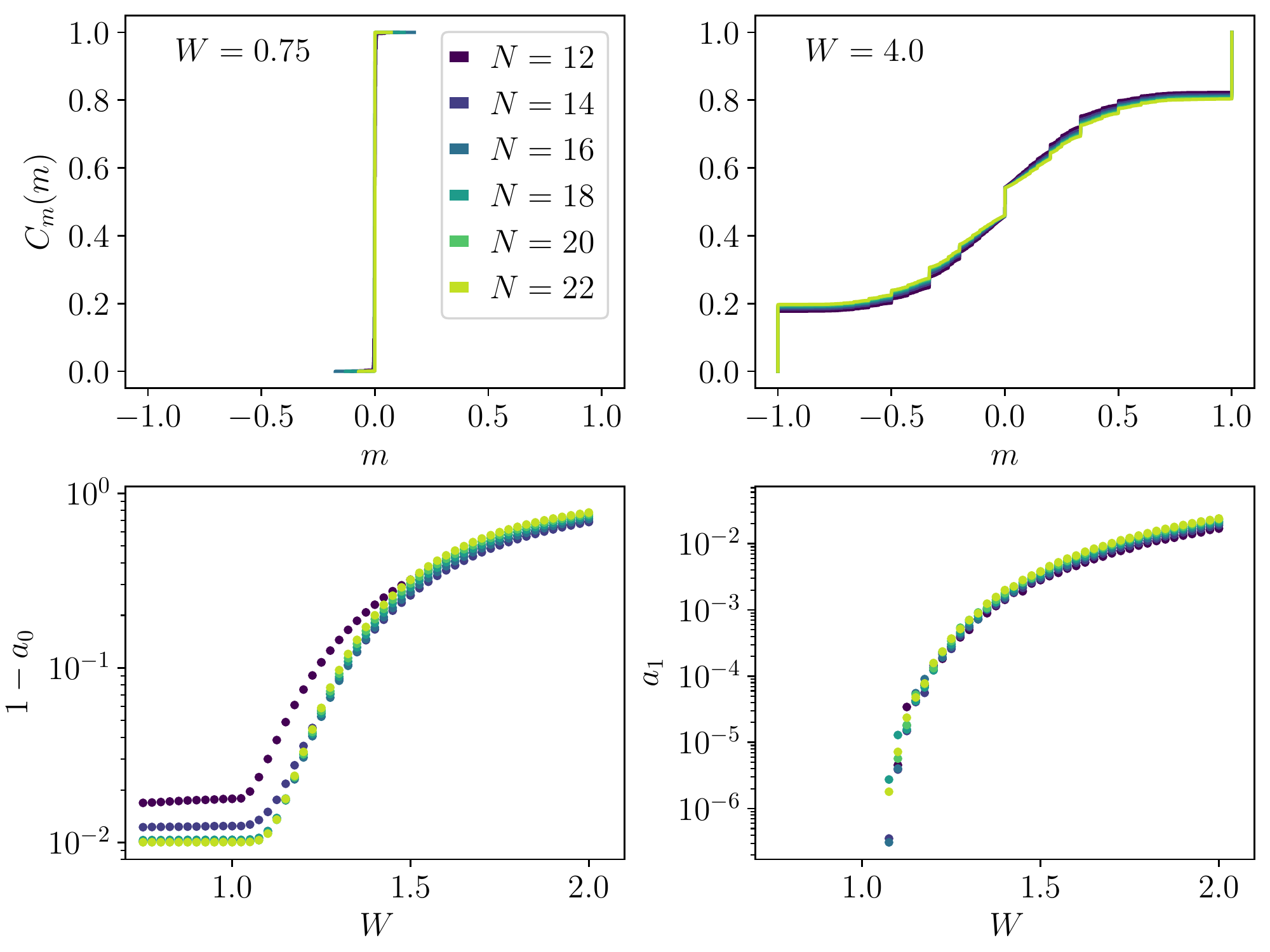}
\caption{Top row: cumulative distribution $C_{m}(m)$ corresponding to $P_m(m)$ shown for two values of $W$, one in the percolating phase and the other in the localised phase. The former has a finite jump only at $m=0$, indicating the presence of a component $\delta(m)$ in $P_m(m)$. In the latter there are additional jumps at $m=\pm1$, indicating the presence of $\delta(\vert m \vert-1)$ components in $P_m(m)$. Bottom row: The deviation of the weights of the $\delta$-functions $a_0$ and $a_1$ (see Eq.~\eqref{eq:pmdelta})
from $1$ and $0$, as a function of disorder $W$, also take on finite values near $W_c\simeq 1$. }
\label{fig:spincumdistribution}
\end{figure}

Note that with the distribution of the $p_I$ 
(Eq.\ \ref{eq:AI})
over the Fock-space considered as analogous to the wavefunction 
probabilities $|\langle\psi |I\rangle|^{2}$ (as in Sec.~\ref{sec:meantypical}), 
$m_\ell$ is 
the counterpart of the expectation value of the operator $\sigma^z_\ell$ in that state, thus drawing a parallel with the quantum system.

Representative numerical results for $P_m(m)$ are shown in Fig.~\ref{fig:spindist}. In the percolating phase ($W<W_c$) the distribution is, as expected, sharply peaked at $m=0$, and sharpens with increasing 
$N$.
As $W$ is increased above the critical disorder, wings in the distribution appear with peaks at $m=\pm1$.
An understanding of the latter can be traced back to the picture that in the localised phase, local spin configurations of finite lengths are frozen out in the cluster.
Hence for all sites within such frozen segments, $\sigma^z_\ell=\pm1$, contributing to these peaks. 
In addition, there also exist spins in the cluster $\mathcal{C}$ which are free to flip, but there is nothing which constrains precisely half of the nodes (Fock-space sites) in this
cluster to have such spins pointing up and the other half pointing down. 
Hence on average they contribute to all values of $m\in (-1,1)$.
Close to the critical point but in the localised phase, there also still exists a finite fraction of spins which can flip under all configurations of their neighbours (see $f_a$ in Fig.~\ref{fig:fraction}) and hence there is also a 
$\delta$-function component at $m=0$.

Showing the histograms as in Fig.~\ref{fig:spindist} with finite bin widths masks somewhat the $\delta$-function 
components of $P_m(m)$ 
in Eq.~\eqref{eq:pmdelta}.
In order to expose that, we also study the cumulative distributions, $C_m(m)=\int_{-1}^{m}dm^\prime P_m(m^\prime)$, shown in Fig.~\ref{fig:spincumdistribution}.
The finite jumps in $C_m(m)$ at $m=0$ and $m=\pm1$ are clear indications of the underlying $\delta$-functions in $P_m(m)$.
In the percolating phase, there is no jump at $m=\pm1$ and there is a clear finite jump at $m=0$ from the $\delta(m)$ component of $P_m(m)$.
On the other hand, in the localised phase, the height of the jump at $m=0$ decreases and finite jumps at $m=\pm1$ appear, indicating that the $\delta(\vert m\vert-1)$ component is finite.
We study the weight of these $\delta$-functions, namely $a_0$ and $a_1$ defined in Eq.~\eqref{eq:pmdelta}, as a function of $W$ by extracting them from the distributions.
As shown in Fig.~\ref{fig:spincumdistribution}(bottom row), the values of $W$ at which $a_0$ and $a_1$ deviate from 1 and 0 are quite close to $W_c$.

\begin{figure}
\includegraphics[width=\linewidth]{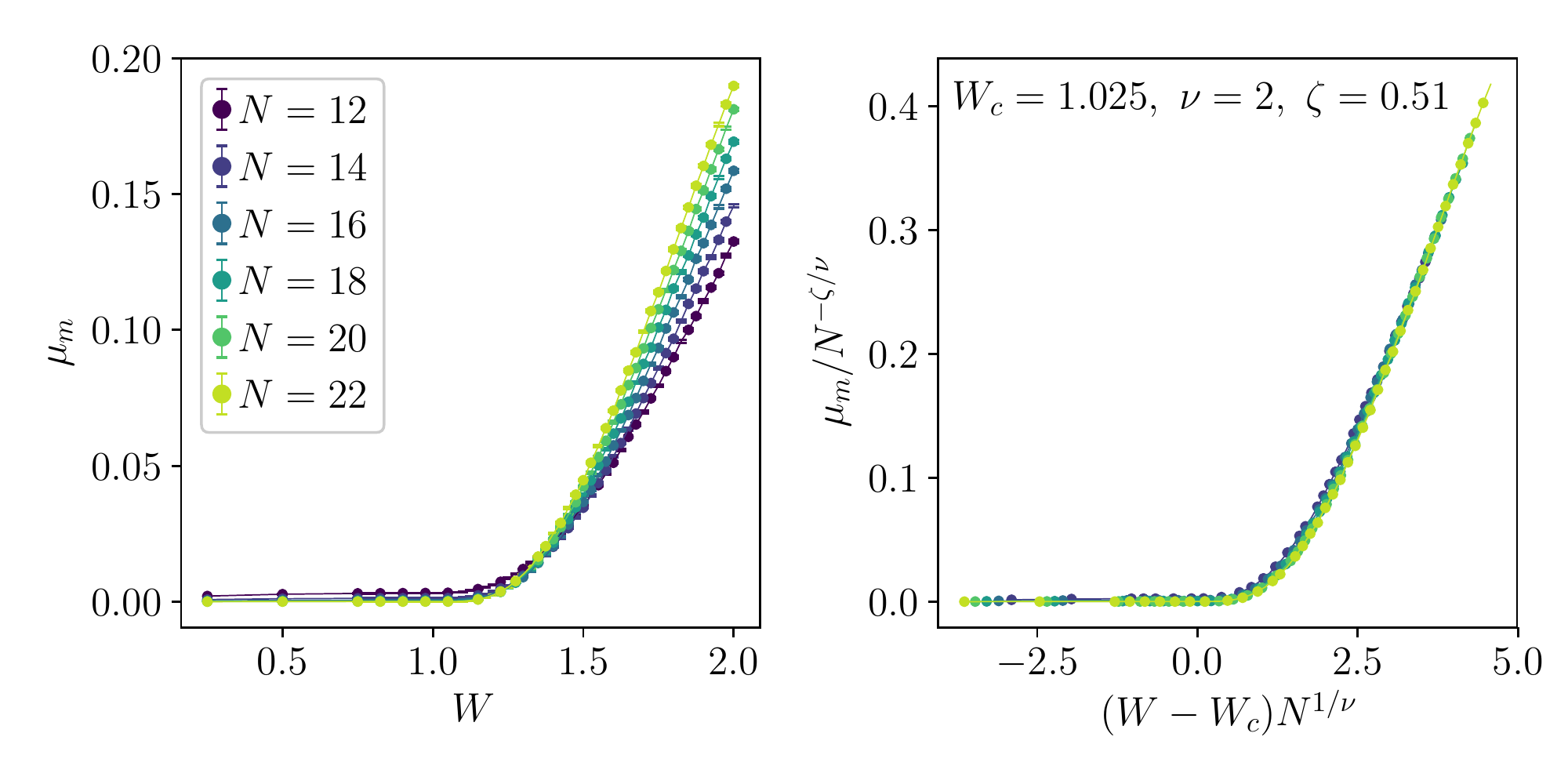}
\caption{Standard deviation of $P_m(m)$, $\mu_m$, shown as a function of disorder $W$ for various system sizes 
$N$. Left panel shows the raw data, whereas the right panel shows data rescaled by $N^{-\zeta/\nu}$ for different system sizes and collapsed onto a common function of $(W-W_c)N^{1/\nu}$. The scaling analysis with $W_c$ and $\nu$ constrained to the values obtained earlier 
yields the exponent $\zeta=0.51$.}
\label{fig:sigma}
\end{figure}

The difference between $P_m(m)$ in the two phases can be made quantitative by studying the variance of the distribution, 
\begin{equation}
\mu_m^2=\int_{-1}^1 P_m(m) \, m^2\, {\rm d}m 
- \bigg(  \int_{-1}^1 P_m(m) \, m \,{\rm d}m \bigg)^2 
\,.
\label{eq:chi-exact}
\end{equation}
In the two extreme limits described above, $W\to0$ and $W\to\infty$, $\mu_m\to0$ and $\mu_m\to1$ respectively.
Moreover, since $\nc=\nh$ in the entire percolating delocalised phase, $\mu_m=0$ throughout this phase. It is non-zero only in the non-percolating localised phase, vanishing as the transition is approached.
Hence $\mu_m$ is also a suitable diagnostic for the transition, and 
the results shown in Fig.~\ref{fig:sigma} 
confirm that. 
In the percolating phase $\mu_m$ is asymptotically vanishing with increasing $N$, whereas in the localised phase it shows a systematic increase with $N$ close to the critical point.
The natural finite-size scaling form for $\mu_m$ is then $\mu_m(N,W)=N^{-\zeta/\nu}g_\mu[(W-W_c)N^{1/\nu}]$. 
As shown in Fig.~\ref{fig:sigma}, constraining $W_c=J/4$ and $\nu=2$, as predicted in Sec.~\ref{sec:physicalxxz}, 
yields a good scaling collapse with the exponent $\zeta=0.51$.

It is significant that the behaviour of the local magnetisation can be used to identify the critical point, giving results that are consistent with the other numerical diagnostics,  
because, as shown in the next section, we can formulate our models as kinetically constrained 
models. With such a formulation local observables are easily accessible to Monte Carlo dynamics,
and hence allow access to much larger system sizes.
In Sec.~\ref{sec:montecarlo} we show that the
autocorrelation function and magnetisation distribution calculated via Monte Carlo dynamics for the TFI model [Eq.~\eqref{eq:hamising}] can be used to locate the transition, and yield values of exponents $\zeta$ and $\nu$ close to that obtained from $\mu_m$ for the random-field XXZ model, illustrating
its universal nature.

\section{Monte Carlo dynamics \label{sec:montecarlo}}

In this section we probe the local magnetisation in the cluster via an exploration of the cluster following Monte Carlo dynamics, where the update rules are intertwined with our Fock-space percolation rules.
The resulting dynamical model is similar to a class of kinetically constrained models known as spin-facilitated Ising models~(see Refs.~\cite{ritort2003glassy,garrahan2011kinetically} for reviews and further references). The dynamical model allows us to compute local observables like those studied in Sec.~\ref{sec:szcluster}, but for much larger system sizes.

The history of kinetically constrained models goes back to theoretical models of glass formers~\cite{fredrickson1984kinetic,fredrickson1985facilitated}, with Markovian dynamics obeying detailed balance with respect to an energy function.
An additional key ingredient is a set of explicit constraints which forbid local transitions between configurations.
Formally, the dynamics of such models can be expressed as a master equation for the probability of the system to be in a configuration (say a spin configuration $\ket{I}$) at time $t$, with the form
\begin{align}
\partial_t p(I,t) = \sum_{{I^\prime}}[&w({I^\prime}\to{I})p({I^\prime},t)-\nonumber\\
&w({I}\to{I^\prime})p(I,t)]\,.
\label{eq:mastereq}
\end{align}
Here $w({I}\to{I^\prime})$ is the transition rate from configuration $\ket{I}$ to $\ket{I^\prime}$, with these two configurations related by a local change of the degrees of freedom -- for example, a single spin flip.
In the absence of a \emph{kinetic constraint}, the rate $w$ is simply a function of the energy difference, $w_0(\mathcal{E}_I-\mathcal{E}_{I^\prime})$
The effect of a kinetic constraint is to forbid some transitions.
It can be expressed in the form
\begin{equation}
w({I}\to{I^\prime}) = \left[\sum_{\ell}\delta_{\bra{I}F_\ell\ket{I^\prime}, 1}\right]w_0(\mathcal{E}_I-\mathcal{E}_{I^\prime}),
\label{eq:constrainedrate}
\end{equation}
where $\{F_\ell\}$ is a set of operators which define the configurations $\ket{I^\prime}$ to which transitions from $\ket{I}$ are allowed.

We now recast the percolation rules defined on the Fock space in Sec.~\ref{sec:percolationmodel} as kinetically constrained dynamics. 
In order to do so, the crucial identification one has to make is that the \emph{active} edges defined via Eq.~\eqref{eq:active} correspond to transitions that are allowed in the kinetically constrained dynamics, and the \emph{inactive} edges to the forbidden ones.
This correspondence can be translated to a dictionary for the two contributions to the rates of transitions as
\begin{equation}
w_0(\mathcal{E}_I-\mathcal{E}_{I^\prime}) = \Theta(J-\vert\mathcal{E}_I-\mathcal{E}_{I^\prime}\vert);
\label{eq:w0}
\end{equation}
and the set of constraints as
\begin{equation}
F_\ell^\pd = \sigma^x_\ell\sigma^x_{\ell+1}+\sigma^y_\ell\sigma^y_{\ell+1}
\label{eq:FlXXZ}
\end{equation}
for the disordered XXZ chain Eq.\ \eqref{eq:hamxxz}, and 
\begin{equation}
F_\ell^\pd = \sigma^x_\ell
\label{eq:FlTFI}
\end{equation}
for the disordered Ising chain Eq.\  \eqref{eq:hamising}.

While the dynamical equation Eq.\ \eqref{eq:mastereq} is written in continuous time, an equivalent discrete-time formulation is much more convenient for simulating the dynamics via Monte Carlo methods.
We implement the dynamics using the following procedure:
\begin{itemize}
	\item We start from a spin configuration $\ket{I_0}$ with its $\mathcal{E}_{I_0}$ close to the mean of the 
	$\mathcal{E}$s.
	\item At any discrete Monte Carlo time, $t_\mathrm{MC} = t$ in the dynamics, let the configuration be $\ket{I}$. One of the constraint operators from the set $\{F_\ell\}$ (Eqs.~\eqref{eq:FlXXZ} and \eqref{eq:FlTFI}) is then randomly chosen and configuration $\ket{I^\prime} = F_\ell\ket{I}$ is identified.
	\item If $w_0(\mathcal{E}_{I^\prime}-\mathcal{E}_I)=1$, the configuration at time $t_\mathrm{MC}=t+1/N$ is $\ket{I^\prime}$, otherwise the configuration at $t+1/N$ remains $\ket{I}$.
	\item A series of $N$ such updates constitutes one Monte Carlo sweep.
\end{itemize}

With this dynamical protocol, what observables will 
reveal the transition from the percolating to the localised phase in their Monte Carlo history?
To answer this question, consider the following simplified space-time pictures for the Monte Carlo dynamics, in the two extreme limits of weak and strong disorder.
For weak disorder, the spins flip easily and the likelihood of a transition event being rejected is low. Hence, the expectation value of a spin averaged over its Monte Carlo history decays quickly. At long enough times, the Monte Carlo average of the spin at each site is mostly homogeneous in space, and the probability distribution over sites of the time-averaged orientation is sharply peaked at zero.
Conversely, for strong disorder, a finite fraction spins do not flip, since $f_n$ [Eq.~\eqref{eq:fan}] is finite.
Hence there are regions in space where the Monte Carlo averaged spin is non-zero. This results in a probability distribution over sites that has weight near $\pm1$ even at long times, as also found by exact enumeration in Fig.~\ref{fig:spindist}.

This suggests that the Monte Carlo average of the spin at each site will have a qualitatively different probability distribution over sites in the percolating and localised phases. The dynamics of its standard deviation is thus an appropriate quantity to study.
The Monte Carlo average of the spin at site $\ell$ is
\begin{equation}
m_\ell^\pd (t_\mathrm{MC}) = \frac{1}{t_\mathrm{MC}+1}\sum_{t=0}^{t_\mathrm{MC}}\bra{I(t)}\sigma^z_\ell\ket{I(t)},
\label{eq:mcaveragelocal}
\end{equation}
where $\ket{I(t)}$ is the spin-configuration at time $t$ in the Monte Carlo trajectory.
Taking statistics over real-space sites and disorder realisations, we obtain the probability distribution of the Monte Carlo averaged local magnetisations denoted by $P_\mathrm{MC}(m,t_\mathrm{MC})$, which naturally depends on 
$t_{\mathrm{MC}}^{\pd}$.
In the limit $t_\mathrm{MC}\to\infty$, the dynamics visits all Fock-space nodes of the cluster $\mathcal{C}$ uniformly, and hence $P_\mathrm{MC}(m,t_\mathrm{MC}\to\infty)\to P_m(m)$ defined in Sec.~\ref{sec:szcluster}.
The dynamics of the variance of $P_\mathrm{MC}(m,t_\mathrm{MC})$,
\begin{align}
\mu^2_\mathrm{MC}(m,t_\mathrm{MC})=&\int_{-1}^1 P_\mathrm{MC}(m,t_\mathrm{MC}) \, m^2\, {\rm d}m 
- \nonumber\\ &\bigg(  \int_{-1}^1 P_\mathrm{MC}(m,t_\mathrm{MC}) \, m \,{\rm d}m \bigg)^2 \,,
\label{eq:mu_montecarlo}
\end{align}
is thus a natural quantity to study. In particular, $\mu_\mathrm{MC}(m,t_\mathrm{MC}\to\infty)\to\mu_m$ 
(defined in Eq.~\eqref{eq:chi-exact}), and hence should analogously act as a diagnostic of the phase transition.

Note that, since we start from a specific initial spin configuration, $P_\mathrm{MC}(m,t_\mathrm{MC}=0)=\delta(\vert m\vert-1)$. Hence the fact that in the localised phase $P_\mathrm{MC}(m,t_\mathrm{MC}\to\infty)$ retains a finite weight on the $\delta$-functions at $m=\pm 1$ is an indication that the Monte Carlo dynamics retains some memory of its initial state, which is typical of a non-ergodic phase.

In the following, we implement the kinetically constrained dynamics for the TFI model Eq.\ \eqref{eq:hamising}, via 
Eqs.~\eqref{eq:w0} and \eqref{eq:FlTFI}.

\begin{figure}
\includegraphics[width=\columnwidth]{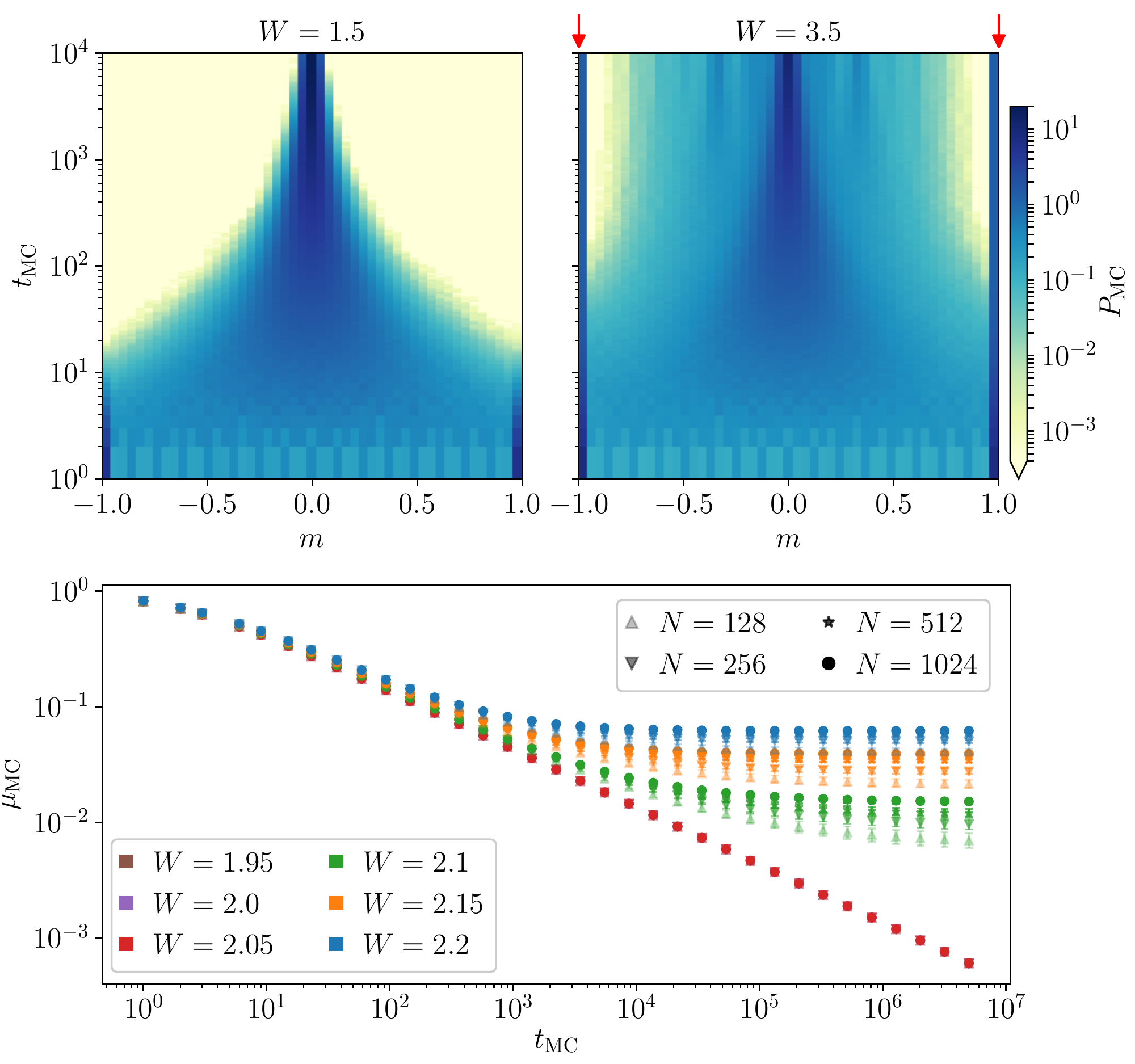}
\caption{Top: For the TFI model, the probability distribution $P_\mathrm{MC}(m,t_\mathrm{MC})$ of the Monte Carlo averaged local spin expectation $m$, Eq.~\eqref{eq:mcaveragelocal}, as a function of Monte Carlo time, $t_\mathrm{MC}$, shown as a colour map. In the percolating phase (left), the distribution becomes unimodal with a peak at $m=0$; whereas in the non-percolating phase (right), the peaks in $P_\mathrm{MC}$ at $m=\pm 1$
persist even at very large $t_\mathrm{MC}$, as indicated via red arrows. 
Bottom: Dynamics of $\mu_\mathrm{MC}(t_\mathrm{MC})$ 
defined in Eq.~\eqref{eq:mu_montecarlo}.
Different colours and symbols represent different values of disorder strength $W$ and system size $N$, respectively. In the ergodic phase, $W<W_c$, $\mu_\mathrm{MC}$ 
shows a decay as $t_\mathrm{MC}^{-1/2}$.
In the non-ergodic phase, $W>W_c$, it saturates to finite values, which grow with increasing $W$. Note that the data for 
$W\le W_c=2.05$ lie on top of each other. Results are obtained over $10^5$ disorder realisations, with statistical errors determined via a standard bootstrap with 500 resamplings.
}
\label{fig:mc_spinsd_dist}
\end{figure}

We first present the results for $P_\mathrm{MC}$ in Fig.~\ref{fig:mc_spinsd_dist}, in the percolating and localised phases.
The distributions at short times are very wide with peaks at $m=\pm1$, since we start from specific classical states.
In the percolating phase (Fig.~\ref{fig:mc_spinsd_dist} top-left panel), the distribution loses weight at finite values of $m$ with increasing $t_\mathrm{MC}$ and becomes a sharply peaked at $m=0$.
In the localised phase by contrast (Fig.~\ref{fig:mc_spinsd_dist} top-right panel), a finite fraction of spins show 
non-ergodic behaviour and stay frozen, resulting in a distribution that is broad with pronounced peaks at $m=\pm1$ even at very large $t_\mathrm{MC}$.

To analyse quantitatively the difference between these distributions,  we study the behaviour of 
$\mu_\mathrm{MC}(t_\mathrm{MC})$ defined in Eq.~\eqref{eq:mu_montecarlo}.
The results are shown in Fig.~\ref{fig:mc_spinsd_dist} (bottom).
For $W<W_c$, $\mu_\mathrm{MC}(t_\mathrm{MC})\sim t_\mathrm{MC}^{-1/2}$. This is simply a consequence of the fact that under Monte Carlo dynamics in the percolating phase, individual spins have rather short correlation times, together with the central limit theorem.
On the other hand, for $W>W_c$, the dynamics is non-ergodic and $\mu_\mathrm{MC}$ saturates to a finite value. 
This is a direct consequence of the fact that in the localised phase, $P_\mathrm{MC}$ retains its broad nature accompanied by peaks at $m=\pm1$.
With increasing $W$, the saturation value of $\mu_\mathrm{MC}$ increases, showing that the system gets more and more localised.

\begin{figure}
\includegraphics[width=\columnwidth]{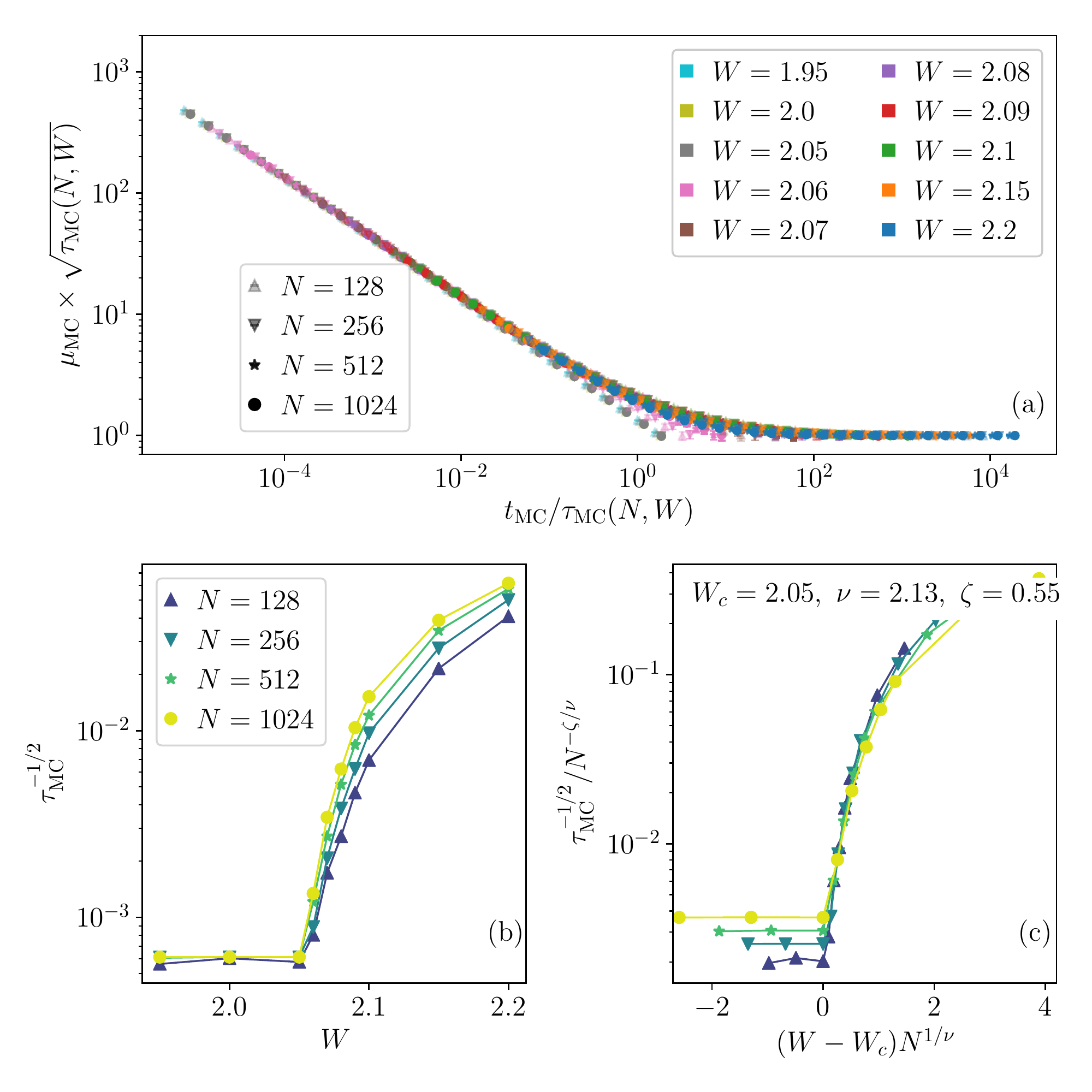}
\caption{(a) Collapse of the data for $\mu_\mathrm{MC}(t_\mathrm{MC})$ for various $N$ and $W$ of the TFI model suggests a scaling form concomitant with that conjectured in Eq.~\eqref{eq:chiscaling}. Different colours and symbols indicate different disorder strengths $W$ and system sizes $N$, respectively. (b) $\tau_\mathrm{MC}^{-1/2}$ as function of $W$ for various $N$, showing that it is a suitable diagnostic of the phase transition. (c) Finite-size scaling analysis of $\tau_\mathrm{MC}^{-1/2}$ shows that the curves for various $N$ can be collapsed onto a common function of $(W-W_c)N^{1/\nu}$ when rescaled with $N^{-\zeta/\nu}$, yielding the displayed values for the critical disorder strength $W_c$ and exponents $\nu$ and $\zeta$.}
\label{fig:mc_tau}
\end{figure}

Having established that $\mu_\mathrm{MC}(t_\mathrm{MC})$ 
distinguishes the two phases, we next analyse its critical properties.
Because $\mu_\mathrm{MC}(t_\mathrm{MC}\to\infty) = \mu_m$, we expect this analysis should give the critical disorder strength and the exponents $\zeta$ and $\nu$.
However, since the limit of $t_\mathrm{MC}\to\infty$ is impossible to achieve in practice, we must perform a scaling analysis taking into account finite $N$ and finite $t_\mathrm{MC}$.

To identify a suitable scaling form, we first note for short  $t_\mathrm{MC}$ that $\mu_\mathrm{MC}(t_\mathrm{MC})$ decays as $t_\mathrm{MC}^{-1/2}$ at essentially all values of $W$. For  $W>W_c$ it saturates at long times at a finite value which depends on both $W$ and $N$.
Inspection of the data in Fig.~\ref{fig:mc_spinsd_dist}(bottom) indicates that there exists a time-scale, $\tau_\mathrm{MC}(N,W)$, at which the $t_\mathrm{MC}^{-1/2}$ decay gives way to saturation, and that this time-scale grows with decreasing $W$, diverging at the critical disorder strength. 
With this in mind, we conjecture the scaling form 
\begin{equation}
\mu_\mathrm{MC}(t_\mathrm{MC}^\pd) = [\tau_\mathrm{MC}(N,W)]^{-1/2} g_\tau\left(\frac{t_\mathrm{MC}}{\tau_\mathrm{MC}}\right)
\label{eq:chiscaling}
\end{equation}
with
\begin{equation}
g_\tau(x)=\begin{cases}
		x^{-1/2}; ~~ x\ll1,\\
		\mathrm{const.} ~~ x\gg1.
	\end{cases}
\end{equation}
This scaling form is ultimately corroborated by collapsing all the curves for $\mu_\mathrm{MC}(t_\mathrm{MC})$ with various $N$ and $W$
in a plot of $\mu_\mathrm{MC}\sqrt{\tau_\mathrm{MC}(N,W)}$ vs $t_\mathrm{MC}/\tau_\mathrm{MC}(N,W)$, 
 as shown in Fig.~\ref{fig:mc_tau}(a).
 
From Eq.~\eqref{eq:chiscaling}, $\tau_\mathrm{MC}(N,W)^{-1/2}$ is proportional to $\mu_m$. It can hence can be used to identify $W_c$ and determine $\zeta$ and $\nu$.
We show $\tau_\mathrm{MC}^{-1/2}$ as function of $W$ for various $N$ in Fig.~\ref{fig:mc_tau}(b). 
Note that the value of $\tau_\mathrm{MC}$ in our simulations is not infinite on the percolating side, but rather set by the largest Monte Carlo time reached.
Nevertheless, a finite-size scaling analysis can be performed by rescaling the data with $N^{-\zeta/\nu}$ and collapsing it onto a common function of $(W-W_c)N^{1/\nu}$, as illustrated in Fig.~\ref{fig:mc_tau}(c).
Such an analysis yields $W_c=2.05$, which is precisely equal to $J/2$ (for $J=4.1$) obtained analytically~\cite{roy2018exact}.
Moreover, the exponents $\nu=2.13$ and $\zeta=0.55$ are remarkably close to those obtained for the disordered XXZ chain via exact enumeration of clusters (see Fig.~\ref{fig:sigma}), indicating the universality of these exponents.

\begin{figure}
\includegraphics[width=\columnwidth]{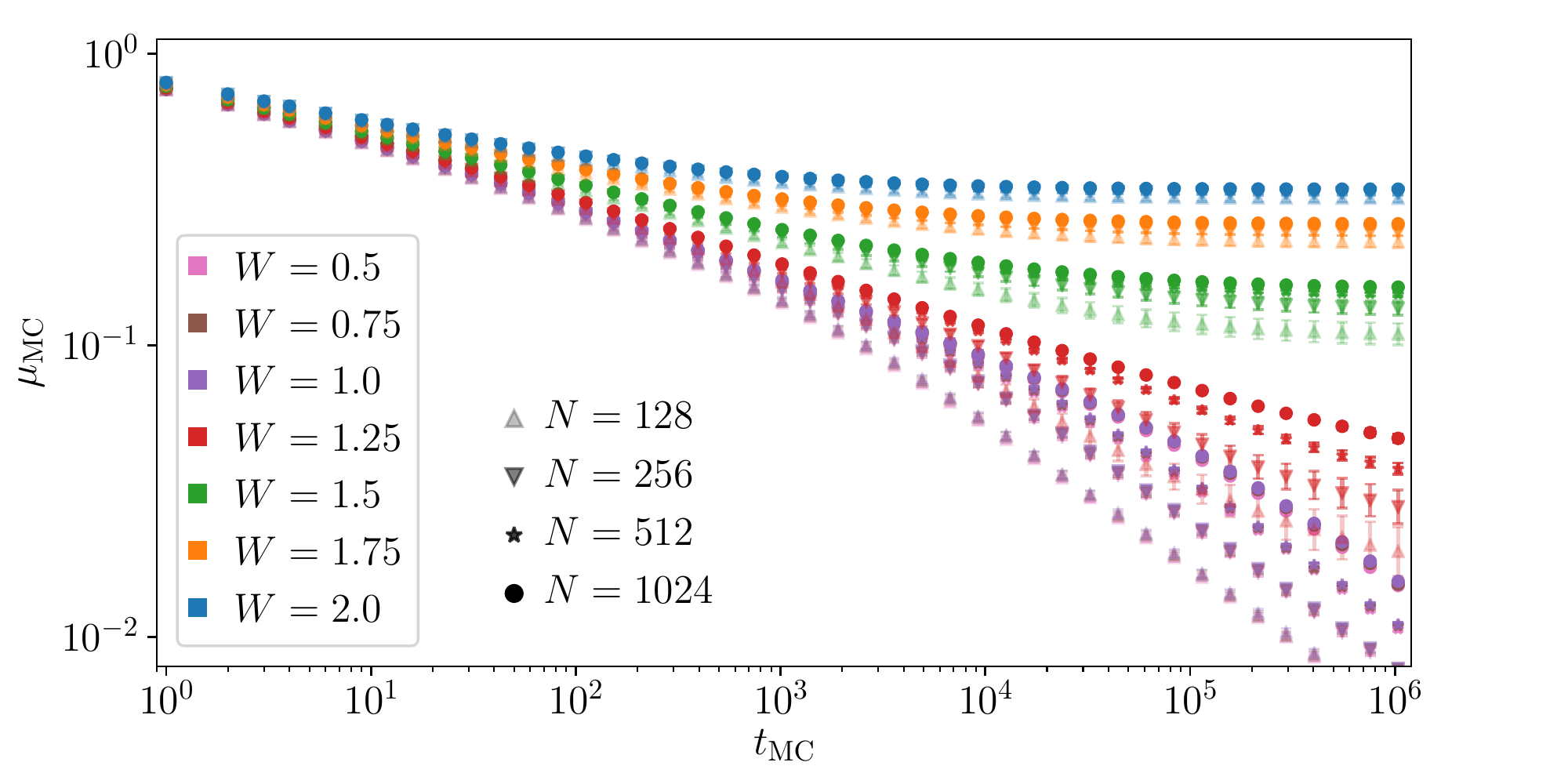}
\caption{Dynamics of $\mu_\mathrm{MC}(t_\mathrm{MC})$ 
for the XXZ model [compare with Fig.~\ref{fig:mc_spinsd_dist} (bottom) for the TFI model].
We use $J_z=1$ and $J=4.1$, for which $W_c=1.025$ for the XXZ model as predicted earlier. Results are obtained from 
$5\times10^4$ disorder realisations, with statistical errors obtained via a standard bootstrap with 500 resamplings.
}
\label{fig:mc_xxz}
\end{figure}

For completeness, we 
add that similar calculations for the disordered XXZ chain yield results consistent with those obtained in Sec.~\ref{sec:numerical}. As an example we show the behaviour of $\mu_\mathrm{MC}(t_\mathrm{MC})$ 
in Fig.~\ref{fig:mc_xxz}. Below and above the critical disorder, $W_c=1.025$, the decay of 
$\mu_\mathrm{MC}(t_\mathrm{MC})$ 
is persistent and arrested respectively. Further details, such as the presence of multiple time scales and decay exponents due to conserved total magnetisation, are left for future work and omitted here in the interests of brevity.

\section{Conclusions and Outlook \label{sec:conclusion}}
In summary, we have studied a classical percolation model in the Fock spaces corresponding to two models of disordered quantum spin chains. In this setting, a percolation transition acts as a proxy for the many-body localisation (MBL) transition in the quantum system.
In Ref.~\cite{roy2018exact} an analytic solution was provided for the critical disorder strength and the correlation length exponent for a quantum transverse field Ising (TFI) spin chain with disordered longitudinal fields.
Here by contrast we have analysed both the random-field XXZ model and the TFI model
in greater detail, to present a physical picture for the phases and the transition, and to corroborate our arguments using a wide variety of numerical diagnostics.
By exactly enumerating the clusters in Fock space for finite systems, we obtained statistics for the cluster sizes, a finite-size scaling of which resulted in critical properties consistent with the analytic results.
In addition, 
connections were made to local observables which, when averaged over the clusters, also carry signatures of the transition.
The set of numerical studies, all of which yield critical properties consistent with each other, is the first 
principal result of the work.

The second 
main result is that dynamics can be introduced by a mapping onto a kinetically constrained model.
Dynamics of such models can be studied via 
well-established methods like Monte Carlo dynamics, allowing access to much larger system sizes.
This is also where the connection to local observables
becomes crucial, since the same local observables,
averaged over the Monte Carlo history, act as diagnostics of the ergodicity-breaking transition, and indeed give critical properties consistent with both the analytical 
and other numerical results.

The classical percolation transition captures certain aspects of the MBL transition remarkably well. We showed for example that the typical cluster size in the percolation problem is directly analogous to the participation entropies of quantum eigenstates. This analogy goes further in terms of the scaling of the typical cluster size, from a volume-law in Fock-space in the percolating phase to a sub-volume law in the localised phase, 
just like
the scaling of participation entropies~\cite{deluca2013ergodicity,luitz2015many}. Fluctuations in the cluster sizes also show a peak at the percolation transition, 
reminiscent of that occurring at the MBL transition in the fluctuations of entanglement entropy (which itself goes from a volume-law to an area-law)~\cite{kjall2014many,luitz2015many,luitz2016long}. Expectation values of local observables also show fundamental differences between many-body localised and ergodic phases, e.g.\ distributions of local magnetisations show a transition from bimodal to unimodal behaviour~\cite{baldwin2016manybody,luitz2016long}. In a very similar fashion, in the classical percolation problem, the local magnetisation averaged over Fock-space sites in the cluster shows a bimodal to unimodal transition.

Despite these clear similarities between the classical percolation transition and the MBL transition, parallels between the two are not yet fully developed.
This leads naturally to the question of what aspects of the quantum transition are not captured by the classical model,
and how the classical model 
might potentially be refined 
to capture them.
For instance, an open question is how 
Griffiths effects, which have been argued
to dominate the physics near the MBL transition~\cite{agarwal2015anomalous,gopalakrishnan2016griffiths,zhang2016many,khemani2017critical}, manifest themselves in such a Fock-space-based approach. 
Relatedly, in our Monte Carlo dynamics we find that the decay of autocorrelations is always diffusive throughout the ergodic phase.
Hence a natural question is what additional constraints might be imposed on the classical model such that it shows a subdiffusive behaviour characteristic of the disordered ergodic phase preceding the MBL transition~\cite{luitz2016anomalous,agarwal2015anomalous,luitz2016extended,luitz2017ergodic}. 
Another delicate issue in the physics of the 
MBL transition is that of mobility edges~\cite{luitz2015many,deroeck2016absence,laumann2014many,welsh2018simple}. In our classical formulation (in common with other approaches to MBL based on resonant percolation in real-space~\cite{potter2015universal,dumitrescu2017scaling}) there is no energy resolution of the cluster
\footnote{The absence of energy resolution is most simply understood by noting that the distribution of cluster sizes is the same when statistics are taken over all clusters from a disorder realisation (which thus includes the entire Fock-space site energy distribution), and when they are taken over only the clusters $\mathcal{C}$ 
`grown' from sites with their site-energies close to the mean of the Fock-space site energy distribution. 
} analogous to eigenenergies. How to incorporate this aspect of the transition within a classical framework is thus a natural issue of interest.

Finally, since the physics of the MBL transition in excited states falls outside the paradigm of equilibrium statistical mechanics, and our percolation model captures certain aspects of it, an interesting open question is how the model can be modified to treat the transitions in manifestly out-of-equilibrium systems, such as Floquet systems and the accompanying anomalous dynamics~\cite{lazarides2015fate,ponte2015many,ponte2015periodically,roy2018anomalous}.

\begin{acknowledgments}
We would like to thank S. Gopalakrishnan, A. Nahum, S. A. Parameswaran and S. Welsh for useful discussions. This work was in part supported by EPSRC Grant No. EP/N01930X/1 and the National Science Foundation under Grant No. NSF PHY-1748958.
\end{acknowledgments}
\bibliography{refs}

\begin{thebibliography}{61}%
\makeatletter
\providecommand \@ifxundefined [1]{%
 \@ifx{#1\undefined}
}%
\providecommand \@ifnum [1]{%
 \ifnum #1\expandafter \@firstoftwo
 \else \expandafter \@secondoftwo
 \fi
}%
\providecommand \@ifx [1]{%
 \ifx #1\expandafter \@firstoftwo
 \else \expandafter \@secondoftwo
 \fi
}%
\providecommand \natexlab [1]{#1}%
\providecommand \enquote  [1]{``#1''}%
\providecommand \bibnamefont  [1]{#1}%
\providecommand \bibfnamefont [1]{#1}%
\providecommand \citenamefont [1]{#1}%
\providecommand \href@noop [0]{\@secondoftwo}%
\providecommand \href [0]{\begingroup \@sanitize@url \@href}%
\providecommand \@href[1]{\@@startlink{#1}\@@href}%
\providecommand \@@href[1]{\endgroup#1\@@endlink}%
\providecommand \@sanitize@url [0]{\catcode `\\12\catcode `\$12\catcode
  `\&12\catcode `\#12\catcode `\^12\catcode `\_12\catcode `\%12\relax}%
\providecommand \@@startlink[1]{}%
\providecommand \@@endlink[0]{}%
\providecommand \url  [0]{\begingroup\@sanitize@url \@url }%
\providecommand \@url [1]{\endgroup\@href {#1}{\urlprefix }}%
\providecommand \urlprefix  [0]{URL }%
\providecommand \Eprint [0]{\href }%
\providecommand \doibase [0]{http://dx.doi.org/}%
\providecommand \selectlanguage [0]{\@gobble}%
\providecommand \bibinfo  [0]{\@secondoftwo}%
\providecommand \bibfield  [0]{\@secondoftwo}%
\providecommand \translation [1]{[#1]}%
\providecommand \BibitemOpen [0]{}%
\providecommand \bibitemStop [0]{}%
\providecommand \bibitemNoStop [0]{.\EOS\space}%
\providecommand \EOS [0]{\spacefactor3000\relax}%
\providecommand \BibitemShut  [1]{\csname bibitem#1\endcsname}%
\let\auto@bib@innerbib\@empty
\bibitem [{\citenamefont {Sachdev}(2011)}]{sachdev2011quantum}%
  \BibitemOpen
  \bibfield  {author} {\bibinfo {author} {\bibfnamefont {S.}~\bibnamefont
  {Sachdev}},\ }\href {https://books.google.co.uk/books?id=F3IkpxwpqSgC} {\emph
  {\bibinfo {title} {Quantum Phase Transitions}}}\ (\bibinfo  {publisher}
  {Cambridge University Press},\ \bibinfo {year} {2011})\BibitemShut {NoStop}%
\bibitem [{\citenamefont {Fisher}(1974)}]{fisher1974renormalization}%
  \BibitemOpen
  \bibfield  {author} {\bibinfo {author} {\bibfnamefont {M.~E.}\ \bibnamefont
  {Fisher}},\ }\bibfield  {title} {\enquote {\bibinfo {title} {The
  renormalization group in the theory of critical behavior},}\ }\href {\doibase
  10.1103/RevModPhys.46.597} {\bibfield  {journal} {\bibinfo  {journal} {Rev.
  Mod. Phys.}\ }\textbf {\bibinfo {volume} {46}},\ \bibinfo {pages} {597}
  (\bibinfo {year} {1974})}\BibitemShut {NoStop}%
\bibitem [{\citenamefont {Wilson}(1975)}]{wilson1975renormalization}%
  \BibitemOpen
  \bibfield  {author} {\bibinfo {author} {\bibfnamefont {K.~G.}\ \bibnamefont
  {Wilson}},\ }\bibfield  {title} {\enquote {\bibinfo {title} {The
  renormalization group: {C}ritical phenomena and the {K}ondo problem},}\
  }\href {\doibase 10.1103/RevModPhys.47.773} {\bibfield  {journal} {\bibinfo
  {journal} {Rev. Mod. Phys.}\ }\textbf {\bibinfo {volume} {47}},\ \bibinfo
  {pages} {773} (\bibinfo {year} {1975})}\BibitemShut {NoStop}%
\bibitem [{\citenamefont {Suzuki}(1976)}]{suzuki1976relationship}%
  \BibitemOpen
  \bibfield  {author} {\bibinfo {author} {\bibfnamefont {M.}~\bibnamefont
  {Suzuki}},\ }\bibfield  {title} {\enquote {\bibinfo {title} {Relationship
  between $d$-dimensional quantal spin systems and $(d+1)$-dimensional ising
  systems: Equivalence, critical exponents and systematic approximants of the
  partition function and spin correlations},}\ }\href
  {https://doi.org/10.1143/PTP.56.1454} {\bibfield  {journal} {\bibinfo
  {journal} {Prog. Theor. Phys.}\ }\textbf {\bibinfo {volume} {56}},\ \bibinfo
  {pages} {1454} (\bibinfo {year} {1976})}\BibitemShut {NoStop}%
\bibitem [{\citenamefont {Pal}\ and\ \citenamefont {Huse}(2010)}]{pal2010many}%
  \BibitemOpen
  \bibfield  {author} {\bibinfo {author} {\bibfnamefont {A.}~\bibnamefont
  {Pal}}\ and\ \bibinfo {author} {\bibfnamefont {D.~A.}\ \bibnamefont {Huse}},\
  }\bibfield  {title} {\enquote {\bibinfo {title} {Many-body localization phase
  transition},}\ }\href {\doibase 10.1103/PhysRevB.82.174411} {\bibfield
  {journal} {\bibinfo  {journal} {Phys. Rev. B}\ }\textbf {\bibinfo {volume}
  {82}},\ \bibinfo {pages} {174411} (\bibinfo {year} {2010})}\BibitemShut
  {NoStop}%
\bibitem [{\citenamefont {Huse}\ \emph {et~al.}(2013)\citenamefont {Huse},
  \citenamefont {Nandkishore}, \citenamefont {Oganesyan}, \citenamefont {Pal},\
  and\ \citenamefont {Sondhi}}]{huse2013localisation}%
  \BibitemOpen
  \bibfield  {author} {\bibinfo {author} {\bibfnamefont {D.~A.}\ \bibnamefont
  {Huse}}, \bibinfo {author} {\bibfnamefont {R.}~\bibnamefont {Nandkishore}},
  \bibinfo {author} {\bibfnamefont {V.}~\bibnamefont {Oganesyan}}, \bibinfo
  {author} {\bibfnamefont {A.}~\bibnamefont {Pal}}, \ and\ \bibinfo {author}
  {\bibfnamefont {S.~L.}\ \bibnamefont {Sondhi}},\ }\bibfield  {title}
  {\enquote {\bibinfo {title} {Localization-protected quantum order},}\ }\href
  {\doibase 10.1103/PhysRevB.88.014206} {\bibfield  {journal} {\bibinfo
  {journal} {Phys. Rev. B}\ }\textbf {\bibinfo {volume} {88}},\ \bibinfo
  {pages} {014206} (\bibinfo {year} {2013})}\BibitemShut {NoStop}%
\bibitem [{\citenamefont {Pekker}\ \emph {et~al.}(2014)\citenamefont {Pekker},
  \citenamefont {Refael}, \citenamefont {Altman}, \citenamefont {Demler},\ and\
  \citenamefont {Oganesyan}}]{pekker2014Hilbert}%
  \BibitemOpen
  \bibfield  {author} {\bibinfo {author} {\bibfnamefont {D.}~\bibnamefont
  {Pekker}}, \bibinfo {author} {\bibfnamefont {G.}~\bibnamefont {Refael}},
  \bibinfo {author} {\bibfnamefont {E.}~\bibnamefont {Altman}}, \bibinfo
  {author} {\bibfnamefont {E.}~\bibnamefont {Demler}}, \ and\ \bibinfo {author}
  {\bibfnamefont {V.}~\bibnamefont {Oganesyan}},\ }\bibfield  {title} {\enquote
  {\bibinfo {title} {Hilbert-glass transition: New universality of
  temperature-tuned many-body dynamical quantum criticality},}\ }\href
  {\doibase 10.1103/PhysRevX.4.011052} {\bibfield  {journal} {\bibinfo
  {journal} {Phys. Rev. X}\ }\textbf {\bibinfo {volume} {4}},\ \bibinfo {pages}
  {011052} (\bibinfo {year} {2014})}\BibitemShut {NoStop}%
\bibitem [{\citenamefont {Parameswaran}\ and\ \citenamefont
  {Vasseur}(2018)}]{paramesawran2018many}%
  \BibitemOpen
  \bibfield  {author} {\bibinfo {author} {\bibfnamefont {S.~A.}\ \bibnamefont
  {Parameswaran}}\ and\ \bibinfo {author} {\bibfnamefont {R.}~\bibnamefont
  {Vasseur}},\ }\bibfield  {title} {\enquote {\bibinfo {title} {Many-body
  localization, symmetry and topology},}\ }\href
  {http://stacks.iop.org/0034-4885/81/i=8/a=082501} {\bibfield  {journal}
  {\bibinfo  {journal} {Rep. Prog. Phys}\ }\textbf {\bibinfo {volume} {81}},\
  \bibinfo {pages} {082501} (\bibinfo {year} {2018})}\BibitemShut {NoStop}%
\bibitem [{\citenamefont {Khemani}\ \emph {et~al.}(2016)\citenamefont
  {Khemani}, \citenamefont {Lazarides}, \citenamefont {Moessner},\ and\
  \citenamefont {Sondhi}}]{khemani2016phase}%
  \BibitemOpen
  \bibfield  {author} {\bibinfo {author} {\bibfnamefont {V.}~\bibnamefont
  {Khemani}}, \bibinfo {author} {\bibfnamefont {A.}~\bibnamefont {Lazarides}},
  \bibinfo {author} {\bibfnamefont {R.}~\bibnamefont {Moessner}}, \ and\
  \bibinfo {author} {\bibfnamefont {S.~L.}\ \bibnamefont {Sondhi}},\ }\bibfield
   {title} {\enquote {\bibinfo {title} {Phase structure of driven quantum
  systems},}\ }\href {\doibase 10.1103/PhysRevLett.116.250401} {\bibfield
  {journal} {\bibinfo  {journal} {Phys. Rev. Lett.}\ }\textbf {\bibinfo
  {volume} {116}},\ \bibinfo {pages} {250401} (\bibinfo {year}
  {2016})}\BibitemShut {NoStop}%
\bibitem [{\citenamefont {Moessner}\ and\ \citenamefont
  {Sondhi}(2017)}]{moessner2017equilibration}%
  \BibitemOpen
  \bibfield  {author} {\bibinfo {author} {\bibfnamefont {R.}~\bibnamefont
  {Moessner}}\ and\ \bibinfo {author} {\bibfnamefont {S.~L.}\ \bibnamefont
  {Sondhi}},\ }\bibfield  {title} {\enquote {\bibinfo {title} {Equilibration
  and order in quantum {F}loquet matter},}\ }\href
  {https://www.nature.com/nphys/journal/v13/n5/abs/nphys4106.html} {\bibfield
  {journal} {\bibinfo  {journal} {Nat. Phys.}\ }\textbf {\bibinfo {volume}
  {13}},\ \bibinfo {pages} {424} (\bibinfo {year} {2017})}\BibitemShut
  {NoStop}%
\bibitem [{\citenamefont {Roy}\ \emph {et~al.}(2018)\citenamefont {Roy},
  \citenamefont {Lazarides}, \citenamefont {Heyl},\ and\ \citenamefont
  {Moessner}}]{roy2018dynamical}%
  \BibitemOpen
  \bibfield  {author} {\bibinfo {author} {\bibfnamefont {S.}~\bibnamefont
  {Roy}}, \bibinfo {author} {\bibfnamefont {A.}~\bibnamefont {Lazarides}},
  \bibinfo {author} {\bibfnamefont {M.}~\bibnamefont {Heyl}}, \ and\ \bibinfo
  {author} {\bibfnamefont {R.}~\bibnamefont {Moessner}},\ }\bibfield  {title}
  {\enquote {\bibinfo {title} {Dynamical potentials for nonequilibrium quantum
  many-body phases},}\ }\href {\doibase 10.1103/PhysRevB.97.205143} {\bibfield
  {journal} {\bibinfo  {journal} {Phys. Rev. B}\ }\textbf {\bibinfo {volume}
  {97}},\ \bibinfo {pages} {205143} (\bibinfo {year} {2018})}\BibitemShut
  {NoStop}%
\bibitem [{\citenamefont {Roy}\ and\ \citenamefont
  {Lazarides}(2018)}]{roy2018nonequilibrium}%
  \BibitemOpen
  \bibfield  {author} {\bibinfo {author} {\bibfnamefont {S.}~\bibnamefont
  {Roy}}\ and\ \bibinfo {author} {\bibfnamefont {A.}~\bibnamefont
  {Lazarides}},\ }\bibfield  {title} {\enquote {\bibinfo {title}
  {Nonequilibrium quantum order at infinite temperature: Spatiotemporal
  correlations and their generating functions},}\ }\href {\doibase
  10.1103/PhysRevB.98.064208} {\bibfield  {journal} {\bibinfo  {journal} {Phys.
  Rev. B}\ }\textbf {\bibinfo {volume} {98}},\ \bibinfo {pages} {064208}
  (\bibinfo {year} {2018})}\BibitemShut {NoStop}%
\bibitem [{\citenamefont {Chan}\ \emph {et~al.}(2018)\citenamefont {Chan},
  \citenamefont {De~Luca},\ and\ \citenamefont {Chalker}}]{chan2018spectral}%
  \BibitemOpen
  \bibfield  {author} {\bibinfo {author} {\bibfnamefont {A.}~\bibnamefont
  {Chan}}, \bibinfo {author} {\bibfnamefont {A.}~\bibnamefont {De~Luca}}, \
  and\ \bibinfo {author} {\bibfnamefont {J.~T.}\ \bibnamefont {Chalker}},\
  }\bibfield  {title} {\enquote {\bibinfo {title} {Spectral statistics in
  spatially extended chaotic quantum many-body systems},}\ }\href {\doibase
  10.1103/PhysRevLett.121.060601} {\bibfield  {journal} {\bibinfo  {journal}
  {Phys. Rev. Lett.}\ }\textbf {\bibinfo {volume} {121}},\ \bibinfo {pages}
  {060601} (\bibinfo {year} {2018})}\BibitemShut {NoStop}%
\bibitem [{\citenamefont {Berdanier}\ \emph {et~al.}(2018)\citenamefont
  {Berdanier}, \citenamefont {Kolodrubetz}, \citenamefont {Parameswaran},\ and\
  \citenamefont {Vasseur}}]{berdanier2018floquet}%
  \BibitemOpen
  \bibfield  {author} {\bibinfo {author} {\bibfnamefont {W.}~\bibnamefont
  {Berdanier}}, \bibinfo {author} {\bibfnamefont {M.}~\bibnamefont
  {Kolodrubetz}}, \bibinfo {author} {\bibfnamefont {S.~A.}\ \bibnamefont
  {Parameswaran}}, \ and\ \bibinfo {author} {\bibfnamefont {R.}~\bibnamefont
  {Vasseur}},\ }\bibfield  {title} {\enquote {\bibinfo {title} {Floquet quantum
  criticality},}\ }\href
  {http://www.pnas.org/content/early/2018/08/28/1805796115} {\bibfield
  {journal} {\bibinfo  {journal} {Proc. Natl. Acad. Sci. U.S.A.}\ } (\bibinfo
  {year} {2018})}\BibitemShut {NoStop}%
\bibitem [{\citenamefont {Gornyi}\ \emph {et~al.}(2005)\citenamefont {Gornyi},
  \citenamefont {Mirlin},\ and\ \citenamefont
  {Polyakov}}]{gornyi2005interacting}%
  \BibitemOpen
  \bibfield  {author} {\bibinfo {author} {\bibfnamefont {I.~V.}\ \bibnamefont
  {Gornyi}}, \bibinfo {author} {\bibfnamefont {A.~D.}\ \bibnamefont {Mirlin}},
  \ and\ \bibinfo {author} {\bibfnamefont {D.~G.}\ \bibnamefont {Polyakov}},\
  }\bibfield  {title} {\enquote {\bibinfo {title} {Interacting electrons in
  disordered wires: Anderson localization and low-${T}$ transport},}\ }\href
  {\doibase 10.1103/PhysRevLett.95.206603} {\bibfield  {journal} {\bibinfo
  {journal} {Phys. Rev. Lett.}\ }\textbf {\bibinfo {volume} {95}},\ \bibinfo
  {pages} {206603} (\bibinfo {year} {2005})}\BibitemShut {NoStop}%
\bibitem [{\citenamefont {Basko}\ \emph {et~al.}(2006)\citenamefont {Basko},
  \citenamefont {Aleiner},\ and\ \citenamefont {Altshuler}}]{basko2006metal}%
  \BibitemOpen
  \bibfield  {author} {\bibinfo {author} {\bibfnamefont {D.~M.}\ \bibnamefont
  {Basko}}, \bibinfo {author} {\bibfnamefont {I.~L.}\ \bibnamefont {Aleiner}},
  \ and\ \bibinfo {author} {\bibfnamefont {B.~L.}\ \bibnamefont {Altshuler}},\
  }\bibfield  {title} {\enquote {\bibinfo {title} {Metal--insulator transition
  in a weakly interacting many-electron system with localized single-particle
  states},}\ }\href
  {http://www.sciencedirect.com/science/article/pii/S0003491605002630}
  {\bibfield  {journal} {\bibinfo  {journal} {Annals of {P}hysics}\ }\textbf
  {\bibinfo {volume} {321}},\ \bibinfo {pages} {1126} (\bibinfo {year}
  {2006})}\BibitemShut {NoStop}%
\bibitem [{\citenamefont {Oganesyan}\ and\ \citenamefont
  {Huse}(2007)}]{oganesyan2007localisation}%
  \BibitemOpen
  \bibfield  {author} {\bibinfo {author} {\bibfnamefont {V.}~\bibnamefont
  {Oganesyan}}\ and\ \bibinfo {author} {\bibfnamefont {D.~A.}\ \bibnamefont
  {Huse}},\ }\bibfield  {title} {\enquote {\bibinfo {title} {Localization of
  interacting fermions at high temperature},}\ }\href {\doibase
  10.1103/PhysRevB.75.155111} {\bibfield  {journal} {\bibinfo  {journal} {Phys.
  Rev. B}\ }\textbf {\bibinfo {volume} {75}},\ \bibinfo {pages} {155111}
  (\bibinfo {year} {2007})}\BibitemShut {NoStop}%
\bibitem [{\citenamefont {\ifmmode \check{Z}\else
  \v{Z}\fi{}nidari\ifmmode~\check{c}\else \v{c}\fi{}}\ \emph
  {et~al.}(2008)\citenamefont {\ifmmode \check{Z}\else
  \v{Z}\fi{}nidari\ifmmode~\check{c}\else \v{c}\fi{}}, \citenamefont {Prosen},\
  and\ \citenamefont {Prelov\ifmmode~\check{s}\else
  \v{s}\fi{}ek}}]{znidaric2008many}%
  \BibitemOpen
  \bibfield  {author} {\bibinfo {author} {\bibfnamefont {M.}~\bibnamefont
  {\ifmmode \check{Z}\else \v{Z}\fi{}nidari\ifmmode~\check{c}\else
  \v{c}\fi{}}}, \bibinfo {author} {\bibfnamefont {T.}~\bibnamefont {Prosen}}, \
  and\ \bibinfo {author} {\bibfnamefont {P.}~\bibnamefont
  {Prelov\ifmmode~\check{s}\else \v{s}\fi{}ek}},\ }\bibfield  {title} {\enquote
  {\bibinfo {title} {Many-body localization in the {H}eisenberg {XXZ} magnet in
  a random field},}\ }\href {\doibase 10.1103/PhysRevB.77.064426} {\bibfield
  {journal} {\bibinfo  {journal} {Phys. Rev. B}\ }\textbf {\bibinfo {volume}
  {77}},\ \bibinfo {pages} {064426} (\bibinfo {year} {2008})}\BibitemShut
  {NoStop}%
\bibitem [{\citenamefont {Kj\"all}\ \emph {et~al.}(2014)\citenamefont
  {Kj\"all}, \citenamefont {Bardarson},\ and\ \citenamefont
  {Pollmann}}]{kjall2014many}%
  \BibitemOpen
  \bibfield  {author} {\bibinfo {author} {\bibfnamefont {J.~A.}\ \bibnamefont
  {Kj\"all}}, \bibinfo {author} {\bibfnamefont {J.~H.}\ \bibnamefont
  {Bardarson}}, \ and\ \bibinfo {author} {\bibfnamefont {F.}~\bibnamefont
  {Pollmann}},\ }\bibfield  {title} {\enquote {\bibinfo {title} {Many-body
  localization in a disordered quantum ising chain},}\ }\href {\doibase
  10.1103/PhysRevLett.113.107204} {\bibfield  {journal} {\bibinfo  {journal}
  {Phys. Rev. Lett.}\ }\textbf {\bibinfo {volume} {113}},\ \bibinfo {pages}
  {107204} (\bibinfo {year} {2014})}\BibitemShut {NoStop}%
\bibitem [{\citenamefont {Laumann}\ \emph {et~al.}(2014)\citenamefont
  {Laumann}, \citenamefont {Pal},\ and\ \citenamefont
  {Scardicchio}}]{laumann2014many}%
  \BibitemOpen
  \bibfield  {author} {\bibinfo {author} {\bibfnamefont {C.~R.}\ \bibnamefont
  {Laumann}}, \bibinfo {author} {\bibfnamefont {A.}~\bibnamefont {Pal}}, \ and\
  \bibinfo {author} {\bibfnamefont {A.}~\bibnamefont {Scardicchio}},\
  }\bibfield  {title} {\enquote {\bibinfo {title} {Many-body mobility edge in a
  mean-field quantum spin glass},}\ }\href {\doibase
  10.1103/PhysRevLett.113.200405} {\bibfield  {journal} {\bibinfo  {journal}
  {Phys. Rev. Lett.}\ }\textbf {\bibinfo {volume} {113}},\ \bibinfo {pages}
  {200405} (\bibinfo {year} {2014})}\BibitemShut {NoStop}%
\bibitem [{\citenamefont {Luitz}\ \emph {et~al.}(2015)\citenamefont {Luitz},
  \citenamefont {Laflorencie},\ and\ \citenamefont {Alet}}]{luitz2015many}%
  \BibitemOpen
  \bibfield  {author} {\bibinfo {author} {\bibfnamefont {D.~J.}\ \bibnamefont
  {Luitz}}, \bibinfo {author} {\bibfnamefont {N.}~\bibnamefont {Laflorencie}},
  \ and\ \bibinfo {author} {\bibfnamefont {F.}~\bibnamefont {Alet}},\
  }\bibfield  {title} {\enquote {\bibinfo {title} {Many-body localization edge
  in the random-field {H}eisenberg chain},}\ }\href {\doibase
  10.1103/PhysRevB.91.081103} {\bibfield  {journal} {\bibinfo  {journal} {Phys.
  Rev. B}\ }\textbf {\bibinfo {volume} {91}},\ \bibinfo {pages} {081103}
  (\bibinfo {year} {2015})}\BibitemShut {NoStop}%
\bibitem [{\citenamefont {Bar~Lev}\ \emph {et~al.}(2015)\citenamefont
  {Bar~Lev}, \citenamefont {Cohen},\ and\ \citenamefont
  {Reichman}}]{lev2015absence}%
  \BibitemOpen
  \bibfield  {author} {\bibinfo {author} {\bibfnamefont {Y.}~\bibnamefont
  {Bar~Lev}}, \bibinfo {author} {\bibfnamefont {G.}~\bibnamefont {Cohen}}, \
  and\ \bibinfo {author} {\bibfnamefont {D.~R.}\ \bibnamefont {Reichman}},\
  }\bibfield  {title} {\enquote {\bibinfo {title} {Absence of diffusion in an
  interacting system of spinless fermions on a one-dimensional disordered
  lattice},}\ }\href {\doibase 10.1103/PhysRevLett.114.100601} {\bibfield
  {journal} {\bibinfo  {journal} {Phys. Rev. Lett.}\ }\textbf {\bibinfo
  {volume} {114}},\ \bibinfo {pages} {100601} (\bibinfo {year}
  {2015})}\BibitemShut {NoStop}%
\bibitem [{\citenamefont {Baldwin}\ \emph {et~al.}(2016)\citenamefont
  {Baldwin}, \citenamefont {Laumann}, \citenamefont {Pal},\ and\ \citenamefont
  {Scardicchio}}]{baldwin2016manybody}%
  \BibitemOpen
  \bibfield  {author} {\bibinfo {author} {\bibfnamefont {C.~L.}\ \bibnamefont
  {Baldwin}}, \bibinfo {author} {\bibfnamefont {C.~R.}\ \bibnamefont
  {Laumann}}, \bibinfo {author} {\bibfnamefont {A.}~\bibnamefont {Pal}}, \ and\
  \bibinfo {author} {\bibfnamefont {A.}~\bibnamefont {Scardicchio}},\
  }\bibfield  {title} {\enquote {\bibinfo {title} {The many-body localized
  phase of the quantum random energy model},}\ }\href {\doibase
  10.1103/PhysRevB.93.024202} {\bibfield  {journal} {\bibinfo  {journal} {Phys.
  Rev. B}\ }\textbf {\bibinfo {volume} {93}},\ \bibinfo {pages} {024202}
  (\bibinfo {year} {2016})}\BibitemShut {NoStop}%
\bibitem [{\citenamefont {Nandkishore}\ and\ \citenamefont
  {Huse}(2015)}]{nandkishore2015many}%
  \BibitemOpen
  \bibfield  {author} {\bibinfo {author} {\bibfnamefont {R.}~\bibnamefont
  {Nandkishore}}\ and\ \bibinfo {author} {\bibfnamefont {D.~A.}\ \bibnamefont
  {Huse}},\ }\bibfield  {title} {\enquote {\bibinfo {title} {Many-body
  localization and thermalization in quantum statistical mechanics},}\ }\href
  {http://www.annualreviews.org/doi/10.1146/annurev-conmatphys-031214-014726}
  {\bibfield  {journal} {\bibinfo  {journal} {Annu. Rev. Condens. Matter
  Phys.}\ }\textbf {\bibinfo {volume} {6}},\ \bibinfo {pages} {15} (\bibinfo
  {year} {2015})}\BibitemShut {NoStop}%
\bibitem [{\citenamefont {Abanin}\ and\ \citenamefont
  {Papi{\'c}}(2017)}]{abanin2017recent}%
  \BibitemOpen
  \bibfield  {author} {\bibinfo {author} {\bibfnamefont {D.~A.}\ \bibnamefont
  {Abanin}}\ and\ \bibinfo {author} {\bibfnamefont {Z.}~\bibnamefont
  {Papi{\'c}}},\ }\bibfield  {title} {\enquote {\bibinfo {title} {Recent
  progress in many-body localization},}\ }\href
  {http://dx.doi.org/10.1002/andp.201700169} {\bibfield  {journal} {\bibinfo
  {journal} {Annalen der Physik}\ }\textbf {\bibinfo {volume} {529}},\ \bibinfo
  {pages} {1700169} (\bibinfo {year} {2017})}\BibitemShut {NoStop}%
\bibitem [{\citenamefont {Alet}\ and\ \citenamefont
  {Laflorencie}(2018)}]{alet2018many}%
  \BibitemOpen
  \bibfield  {author} {\bibinfo {author} {\bibfnamefont {F.}~\bibnamefont
  {Alet}}\ and\ \bibinfo {author} {\bibfnamefont {N.}~\bibnamefont
  {Laflorencie}},\ }\bibfield  {title} {\enquote {\bibinfo {title} {Many-body
  localization: an introduction and selected topics},}\ }\href
  {https://www.sciencedirect.com/science/article/pii/S163107051830032X?via%3Dihub}
  {\bibfield  {journal} {\bibinfo  {journal} {Comptes Rendus Physique}\ }
  (\bibinfo {year} {2018})}\BibitemShut {NoStop}%
\bibitem [{\citenamefont {Potter}\ \emph {et~al.}(2015)\citenamefont {Potter},
  \citenamefont {Vasseur},\ and\ \citenamefont
  {Parameswaran}}]{potter2015universal}%
  \BibitemOpen
  \bibfield  {author} {\bibinfo {author} {\bibfnamefont {A.~C.}\ \bibnamefont
  {Potter}}, \bibinfo {author} {\bibfnamefont {R.}~\bibnamefont {Vasseur}}, \
  and\ \bibinfo {author} {\bibfnamefont {S.~A.}\ \bibnamefont {Parameswaran}},\
  }\bibfield  {title} {\enquote {\bibinfo {title} {Universal properties of
  many-body delocalization transitions},}\ }\href {\doibase
  10.1103/PhysRevX.5.031033} {\bibfield  {journal} {\bibinfo  {journal} {Phys.
  Rev. X}\ }\textbf {\bibinfo {volume} {5}},\ \bibinfo {pages} {031033}
  (\bibinfo {year} {2015})}\BibitemShut {NoStop}%
\bibitem [{\citenamefont {Vosk}\ \emph {et~al.}(2015)\citenamefont {Vosk},
  \citenamefont {Huse},\ and\ \citenamefont {Altman}}]{vosk2015theory}%
  \BibitemOpen
  \bibfield  {author} {\bibinfo {author} {\bibfnamefont {R.}~\bibnamefont
  {Vosk}}, \bibinfo {author} {\bibfnamefont {D.~A.}\ \bibnamefont {Huse}}, \
  and\ \bibinfo {author} {\bibfnamefont {E.}~\bibnamefont {Altman}},\
  }\bibfield  {title} {\enquote {\bibinfo {title} {Theory of the many-body
  localization transition in one-dimensional systems},}\ }\href {\doibase
  10.1103/PhysRevX.5.031032} {\bibfield  {journal} {\bibinfo  {journal} {Phys.
  Rev. X}\ }\textbf {\bibinfo {volume} {5}},\ \bibinfo {pages} {031032}
  (\bibinfo {year} {2015})}\BibitemShut {NoStop}%
\bibitem [{\citenamefont {Dumitrescu}\ \emph {et~al.}(2017)\citenamefont
  {Dumitrescu}, \citenamefont {Vasseur},\ and\ \citenamefont
  {Potter}}]{dumitrescu2017scaling}%
  \BibitemOpen
  \bibfield  {author} {\bibinfo {author} {\bibfnamefont {P.~T.}\ \bibnamefont
  {Dumitrescu}}, \bibinfo {author} {\bibfnamefont {R.}~\bibnamefont {Vasseur}},
  \ and\ \bibinfo {author} {\bibfnamefont {A.~C.}\ \bibnamefont {Potter}},\
  }\bibfield  {title} {\enquote {\bibinfo {title} {Scaling theory of
  entanglement at the many-body localization transition},}\ }\href {\doibase
  10.1103/PhysRevLett.119.110604} {\bibfield  {journal} {\bibinfo  {journal}
  {Phys. Rev. Lett.}\ }\textbf {\bibinfo {volume} {119}},\ \bibinfo {pages}
  {110604} (\bibinfo {year} {2017})}\BibitemShut {NoStop}%
\bibitem [{\citenamefont {Goremykina}\ \emph {et~al.}(2018)\citenamefont
  {Goremykina}, \citenamefont {Vasseur},\ and\ \citenamefont
  {Serbyn}}]{goremykina2018analytically}%
  \BibitemOpen
  \bibfield  {author} {\bibinfo {author} {\bibfnamefont {A.}~\bibnamefont
  {Goremykina}}, \bibinfo {author} {\bibfnamefont {R.}~\bibnamefont {Vasseur}},
  \ and\ \bibinfo {author} {\bibfnamefont {M.}~\bibnamefont {Serbyn}},\
  }\bibfield  {title} {\enquote {\bibinfo {title} {Analytically solvable
  renormalization group for the many-body localization transition},}\ }\href
  {https://arxiv.org/abs/1807.04285} {\bibfield  {journal} {\bibinfo  {journal}
  {arXiv:1807.04285}\ } (\bibinfo {year} {2018})}\BibitemShut {NoStop}%
\bibitem [{\citenamefont {Dumitrescu}\ \emph {et~al.}(2018)\citenamefont
  {Dumitrescu}, \citenamefont {Parameswaran}, \citenamefont {Goremykina},
  \citenamefont {Serbyn},\ and\ \citenamefont
  {Vasseur}}]{dumitrescu2018kosterlitz}%
  \BibitemOpen
  \bibfield  {author} {\bibinfo {author} {\bibfnamefont {P.~T.}\ \bibnamefont
  {Dumitrescu}}, \bibinfo {author} {\bibfnamefont {S.~A.}\ \bibnamefont
  {Parameswaran}}, \bibinfo {author} {\bibfnamefont {A.}~\bibnamefont
  {Goremykina}}, \bibinfo {author} {\bibfnamefont {M.}~\bibnamefont {Serbyn}},
  \ and\ \bibinfo {author} {\bibfnamefont {R.}~\bibnamefont {Vasseur}},\
  }\bibfield  {title} {\enquote {\bibinfo {title} {Kosterlitz-{T}houless
  scaling at many-body localization phase transitions},}\ }\href
  {https://arxiv.org/abs/1811.03103} {\bibfield  {journal} {\bibinfo  {journal}
  {arXiv:1811.03103}\ } (\bibinfo {year} {2018})}\BibitemShut {NoStop}%
\bibitem [{\citenamefont {{Roy}}\ \emph {et~al.}(2018)\citenamefont {{Roy}},
  \citenamefont {{Logan}},\ and\ \citenamefont {{Chalker}}}]{roy2018exact}%
  \BibitemOpen
  \bibfield  {author} {\bibinfo {author} {\bibfnamefont {S.}~\bibnamefont
  {{Roy}}}, \bibinfo {author} {\bibfnamefont {D.~E.}\ \bibnamefont {{Logan}}},
  \ and\ \bibinfo {author} {\bibfnamefont {J.~T.}\ \bibnamefont {{Chalker}}},\
  }\bibfield  {title} {\enquote {\bibinfo {title} {{Exact solution of a
  percolation analogue for the many-body localisation transition}},}\ }\href
  {https://arxiv.org/abs/1812.05115} {\bibfield  {journal} {\bibinfo  {journal}
  {arXiv:1812.05115}\ } (\bibinfo {year} {2018})}\BibitemShut {NoStop}%
\bibitem [{\citenamefont {Logan}\ and\ \citenamefont
  {Welsh}(2018)}]{logan2018many}%
  \BibitemOpen
  \bibfield  {author} {\bibinfo {author} {\bibfnamefont {D.~E.}\ \bibnamefont
  {Logan}}\ and\ \bibinfo {author} {\bibfnamefont {S.}~\bibnamefont {Welsh}},\
  }\bibfield  {title} {\enquote {\bibinfo {title} {Many-body localization in
  {F}ock-space: a local perspective},}\ }\href
  {https://arxiv.org/abs/1806.01688} {\bibfield  {journal} {\bibinfo  {journal}
  {arXiv:1806.01688}\ } (\bibinfo {year} {2018})}\BibitemShut {NoStop}%
\bibitem [{\citenamefont {Ritort}\ and\ \citenamefont
  {Sollich}(2003)}]{ritort2003glassy}%
  \BibitemOpen
  \bibfield  {author} {\bibinfo {author} {\bibfnamefont {F.}~\bibnamefont
  {Ritort}}\ and\ \bibinfo {author} {\bibfnamefont {P.}~\bibnamefont
  {Sollich}},\ }\bibfield  {title} {\enquote {\bibinfo {title} {Glassy dynamics
  of kinetically constrained models},}\ }\href
  {https://www.tandfonline.com/doi/abs/10.1080/0001873031000093582} {\bibfield
  {journal} {\bibinfo  {journal} {Advances in Physics}\ }\textbf {\bibinfo
  {volume} {52}},\ \bibinfo {pages} {219--342} (\bibinfo {year}
  {2003})}\BibitemShut {NoStop}%
\bibitem [{\citenamefont {Garrahan}\ \emph {et~al.}(2011)\citenamefont
  {Garrahan}, \citenamefont {Sollich},\ and\ \citenamefont
  {Toninelli}}]{garrahan2011kinetically}%
  \BibitemOpen
  \bibfield  {author} {\bibinfo {author} {\bibfnamefont {J.~P.}\ \bibnamefont
  {Garrahan}}, \bibinfo {author} {\bibfnamefont {P.}~\bibnamefont {Sollich}}, \
  and\ \bibinfo {author} {\bibfnamefont {C.}~\bibnamefont {Toninelli}},\
  }\bibfield  {title} {\enquote {\bibinfo {title} {Kinetically constrained
  models},}\ }in\ \href
  {http://www.oxfordscholarship.com/view/10.1093/acprof:oso/9780199691470.001.0001/acprof-9780199691470-chapter-10}
  {\emph {\bibinfo {booktitle} {Dynamical heterogeneities in glasses, colloids,
  and granular media}}},\ \bibinfo {editor} {edited by\ \bibinfo {editor}
  {\bibfnamefont {L.}~\bibnamefont {Berthier}}, \bibinfo {editor}
  {\bibfnamefont {G.}~\bibnamefont {Biroli}}, \bibinfo {editor} {\bibfnamefont
  {J.-P.}\ \bibnamefont {Bouchaud}}, \bibinfo {editor} {\bibfnamefont
  {L.}~\bibnamefont {Cipelletti}}, \ and\ \bibinfo {editor} {\bibfnamefont
  {W.}~\bibnamefont {van Saarloos}}}\ (\bibinfo  {publisher} {Oxford University
  Press},\ \bibinfo {address} {Oxford},\ \bibinfo {year} {2011})\BibitemShut
  {NoStop}%
\bibitem [{\citenamefont {Fredrickson}\ and\ \citenamefont
  {Andersen}(1984)}]{fredrickson1984kinetic}%
  \BibitemOpen
  \bibfield  {author} {\bibinfo {author} {\bibfnamefont {G.~H.}\ \bibnamefont
  {Fredrickson}}\ and\ \bibinfo {author} {\bibfnamefont {H.~C.}\ \bibnamefont
  {Andersen}},\ }\bibfield  {title} {\enquote {\bibinfo {title} {Kinetic ising
  model of the glass transition},}\ }\href {\doibase
  10.1103/PhysRevLett.53.1244} {\bibfield  {journal} {\bibinfo  {journal}
  {Phys. Rev. Lett.}\ }\textbf {\bibinfo {volume} {53}},\ \bibinfo {pages}
  {1244--1247} (\bibinfo {year} {1984})}\BibitemShut {NoStop}%
\bibitem [{\citenamefont {Fredrickson}\ and\ \citenamefont
  {Andersen}(1985)}]{fredrickson1985facilitated}%
  \BibitemOpen
  \bibfield  {author} {\bibinfo {author} {\bibfnamefont {G.~H.}\ \bibnamefont
  {Fredrickson}}\ and\ \bibinfo {author} {\bibfnamefont {H.~C}\ \bibnamefont
  {Andersen}},\ }\bibfield  {title} {\enquote {\bibinfo {title} {Facilitated
  kinetic ising models and the glass transition},}\ }\href
  {https://aip.scitation.org/doi/10.1063/1.449662} {\bibfield  {journal}
  {\bibinfo  {journal} {J. Chem. Phys.}\ }\textbf {\bibinfo {volume} {83}},\
  \bibinfo {pages} {5822--5831} (\bibinfo {year} {1985})}\BibitemShut {NoStop}%
\bibitem [{\citenamefont {van Horssen}\ \emph {et~al.}(2015)\citenamefont {van
  Horssen}, \citenamefont {Levi},\ and\ \citenamefont
  {Garrahan}}]{horssen2015dynamics}%
  \BibitemOpen
  \bibfield  {author} {\bibinfo {author} {\bibfnamefont {M.}~\bibnamefont {van
  Horssen}}, \bibinfo {author} {\bibfnamefont {E.}~\bibnamefont {Levi}}, \ and\
  \bibinfo {author} {\bibfnamefont {J.~P.}\ \bibnamefont {Garrahan}},\
  }\bibfield  {title} {\enquote {\bibinfo {title} {Dynamics of many-body
  localization in a translation-invariant quantum glass model},}\ }\href
  {\doibase 10.1103/PhysRevB.92.100305} {\bibfield  {journal} {\bibinfo
  {journal} {Phys. Rev. B}\ }\textbf {\bibinfo {volume} {92}},\ \bibinfo
  {pages} {100305} (\bibinfo {year} {2015})}\BibitemShut {NoStop}%
\bibitem [{\citenamefont {Hickey}\ \emph {et~al.}(2016)\citenamefont {Hickey},
  \citenamefont {Genway},\ and\ \citenamefont
  {Garrahan}}]{hickey2016signatures}%
  \BibitemOpen
  \bibfield  {author} {\bibinfo {author} {\bibfnamefont {J.~M.}\ \bibnamefont
  {Hickey}}, \bibinfo {author} {\bibfnamefont {S.}~\bibnamefont {Genway}}, \
  and\ \bibinfo {author} {\bibfnamefont {J.~P.}\ \bibnamefont {Garrahan}},\
  }\bibfield  {title} {\enquote {\bibinfo {title} {Signatures of many-body
  localisation in a system without disorder and the relation to a glass
  transition},}\ }\href
  {http://iopscience.iop.org/article/10.1088/1742-5468/2016/05/054047/meta}
  {\bibfield  {journal} {\bibinfo  {journal} {J. Stat. Mech.}\ ,\ \bibinfo
  {pages} {054047}} (\bibinfo {year} {2016})}\BibitemShut {NoStop}%
\bibitem [{\citenamefont {Lan}\ \emph {et~al.}(2018)\citenamefont {Lan},
  \citenamefont {van Horssen}, \citenamefont {Powell},\ and\ \citenamefont
  {Garrahan}}]{lan2018quantum}%
  \BibitemOpen
  \bibfield  {author} {\bibinfo {author} {\bibfnamefont {Z.}~\bibnamefont
  {Lan}}, \bibinfo {author} {\bibfnamefont {M.}~\bibnamefont {van Horssen}},
  \bibinfo {author} {\bibfnamefont {S.}~\bibnamefont {Powell}}, \ and\ \bibinfo
  {author} {\bibfnamefont {J.~P.}\ \bibnamefont {Garrahan}},\ }\bibfield
  {title} {\enquote {\bibinfo {title} {Quantum slow relaxation and
  metastability due to dynamical constraints},}\ }\href {\doibase
  10.1103/PhysRevLett.121.040603} {\bibfield  {journal} {\bibinfo  {journal}
  {Phys. Rev. Lett.}\ }\textbf {\bibinfo {volume} {121}},\ \bibinfo {pages}
  {040603} (\bibinfo {year} {2018})}\BibitemShut {NoStop}%
\bibitem [{\citenamefont {Welsh}\ and\ \citenamefont
  {Logan}(2018)}]{welsh2018simple}%
  \BibitemOpen
  \bibfield  {author} {\bibinfo {author} {\bibfnamefont {S.}~\bibnamefont
  {Welsh}}\ and\ \bibinfo {author} {\bibfnamefont {D.~E.}\ \bibnamefont
  {Logan}},\ }\bibfield  {title} {\enquote {\bibinfo {title} {Simple
  probability distributions on a fock-space lattice},}\ }\href
  {http://stacks.iop.org/0953-8984/30/i=40/a=405601} {\bibfield  {journal}
  {\bibinfo  {journal} {J. Phys.: Condens. Matter}\ }\textbf {\bibinfo {volume}
  {30}},\ \bibinfo {pages} {405601} (\bibinfo {year} {2018})}\BibitemShut
  {NoStop}%
\bibitem [{\citenamefont {Imbrie}(2016)}]{imbrie2016many}%
  \BibitemOpen
  \bibfield  {author} {\bibinfo {author} {\bibfnamefont {J.~Z.}\ \bibnamefont
  {Imbrie}},\ }\bibfield  {title} {\enquote {\bibinfo {title} {On many-body
  localization for quantum spin chains},}\ }\href
  {https://link.springer.com/article/10.1007%2Fs10955-016-1508-x} {\bibfield
  {journal} {\bibinfo  {journal} {J. Stat. Phys.}\ }\textbf {\bibinfo {volume}
  {163}},\ \bibinfo {pages} {998} (\bibinfo {year} {2016})}\BibitemShut
  {NoStop}%
\bibitem [{\citenamefont {Stauffer}\ and\ \citenamefont
  {Aharony}(1994)}]{stauffer2014introduction}%
  \BibitemOpen
  \bibfield  {author} {\bibinfo {author} {\bibfnamefont {D.}~\bibnamefont
  {Stauffer}}\ and\ \bibinfo {author} {\bibfnamefont {A.}~\bibnamefont
  {Aharony}},\ }\href
  {https://www.crcpress.com/Introduction-To-Percolation-Theory-Revised-Second-Edition/Stauffer-Aharony/p/book/9780748402533}
  {\emph {\bibinfo {title} {Introduction to percolation theory: revised second
  edition}}}\ (\bibinfo  {publisher} {CRC press},\ \bibinfo {year}
  {1994})\BibitemShut {NoStop}%
\bibitem [{\citenamefont {De~Luca}\ and\ \citenamefont
  {Scardicchio}(2013)}]{deluca2013ergodicity}%
  \BibitemOpen
  \bibfield  {author} {\bibinfo {author} {\bibfnamefont {A.}~\bibnamefont
  {De~Luca}}\ and\ \bibinfo {author} {\bibfnamefont {A.}~\bibnamefont
  {Scardicchio}},\ }\bibfield  {title} {\enquote {\bibinfo {title} {Ergodicity
  breaking in a model showing many-body localization},}\ }\href
  {http://iopscience.iop.org/article/10.1209/0295-5075/101/37003/meta}
  {\bibfield  {journal} {\bibinfo  {journal} {Europhys. Lett.}\ }\textbf
  {\bibinfo {volume} {101}},\ \bibinfo {pages} {37003} (\bibinfo {year}
  {2013})}\BibitemShut {NoStop}%
\bibitem [{\citenamefont {Harris}(1974)}]{harris1974effect}%
  \BibitemOpen
  \bibfield  {author} {\bibinfo {author} {\bibfnamefont {A.~B.}\ \bibnamefont
  {Harris}},\ }\bibfield  {title} {\enquote {\bibinfo {title} {Effect of random
  defects on the critical behaviour of ising models},}\ }\href
  {http://stacks.iop.org/0022-3719/7/i=9/a=009} {\bibfield  {journal} {\bibinfo
   {journal} {J. Phys. C: Solid State Physics}\ }\textbf {\bibinfo {volume}
  {7}},\ \bibinfo {pages} {1671} (\bibinfo {year} {1974})}\BibitemShut
  {NoStop}%
\bibitem [{\citenamefont {Chayes}\ \emph {et~al.}(1986)\citenamefont {Chayes},
  \citenamefont {Chayes}, \citenamefont {Fisher},\ and\ \citenamefont
  {Spencer}}]{chayes1986finite}%
  \BibitemOpen
  \bibfield  {author} {\bibinfo {author} {\bibfnamefont {J.~T.}\ \bibnamefont
  {Chayes}}, \bibinfo {author} {\bibfnamefont {L.}~\bibnamefont {Chayes}},
  \bibinfo {author} {\bibfnamefont {D.~S.}\ \bibnamefont {Fisher}}, \ and\
  \bibinfo {author} {\bibfnamefont {T.}~\bibnamefont {Spencer}},\ }\bibfield
  {title} {\enquote {\bibinfo {title} {Finite-size scaling and correlation
  lengths for disordered systems},}\ }\href {\doibase
  10.1103/PhysRevLett.57.2999} {\bibfield  {journal} {\bibinfo  {journal}
  {Phys. Rev. Lett.}\ }\textbf {\bibinfo {volume} {57}},\ \bibinfo {pages}
  {2999--3002} (\bibinfo {year} {1986})}\BibitemShut {NoStop}%
\bibitem [{\citenamefont {Chandran}\ \emph {et~al.}(2015)\citenamefont
  {Chandran}, \citenamefont {Laumann},\ and\ \citenamefont
  {Oganesyan}}]{chandran2015finite}%
  \BibitemOpen
  \bibfield  {author} {\bibinfo {author} {\bibfnamefont {A.}~\bibnamefont
  {Chandran}}, \bibinfo {author} {\bibfnamefont {C.~R.}\ \bibnamefont
  {Laumann}}, \ and\ \bibinfo {author} {\bibfnamefont {V.}~\bibnamefont
  {Oganesyan}},\ }\bibfield  {title} {\enquote {\bibinfo {title} {Finite size
  scaling bounds on many-body localized phase transitions},}\ }\href
  {https://arxiv.org/abs/1509.04285} {\bibfield  {journal} {\bibinfo  {journal}
  {arXiv:1509.04285}\ } (\bibinfo {year} {2015})}\BibitemShut {NoStop}%
\bibitem [{\citenamefont {Luitz}(2016)}]{luitz2016long}%
  \BibitemOpen
  \bibfield  {author} {\bibinfo {author} {\bibfnamefont {D.~J.}\ \bibnamefont
  {Luitz}},\ }\bibfield  {title} {\enquote {\bibinfo {title} {Long tail
  distributions near the many-body localization transition},}\ }\href {\doibase
  10.1103/PhysRevB.93.134201} {\bibfield  {journal} {\bibinfo  {journal} {Phys.
  Rev. B}\ }\textbf {\bibinfo {volume} {93}},\ \bibinfo {pages} {134201}
  (\bibinfo {year} {2016})}\BibitemShut {NoStop}%
\bibitem [{\citenamefont {Agarwal}\ \emph {et~al.}(2015)\citenamefont
  {Agarwal}, \citenamefont {Gopalakrishnan}, \citenamefont {Knap},
  \citenamefont {M\"uller},\ and\ \citenamefont
  {Demler}}]{agarwal2015anomalous}%
  \BibitemOpen
  \bibfield  {author} {\bibinfo {author} {\bibfnamefont {K.}~\bibnamefont
  {Agarwal}}, \bibinfo {author} {\bibfnamefont {S.}~\bibnamefont
  {Gopalakrishnan}}, \bibinfo {author} {\bibfnamefont {M.}~\bibnamefont
  {Knap}}, \bibinfo {author} {\bibfnamefont {M.}~\bibnamefont {M\"uller}}, \
  and\ \bibinfo {author} {\bibfnamefont {E.}~\bibnamefont {Demler}},\
  }\bibfield  {title} {\enquote {\bibinfo {title} {Anomalous diffusion and
  griffiths effects near the many-body localization transition},}\ }\href
  {\doibase 10.1103/PhysRevLett.114.160401} {\bibfield  {journal} {\bibinfo
  {journal} {Phys. Rev. Lett.}\ }\textbf {\bibinfo {volume} {114}},\ \bibinfo
  {pages} {160401} (\bibinfo {year} {2015})}\BibitemShut {NoStop}%
\bibitem [{\citenamefont {Gopalakrishnan}\ \emph {et~al.}(2016)\citenamefont
  {Gopalakrishnan}, \citenamefont {Agarwal}, \citenamefont {Demler},
  \citenamefont {Huse},\ and\ \citenamefont
  {Knap}}]{gopalakrishnan2016griffiths}%
  \BibitemOpen
  \bibfield  {author} {\bibinfo {author} {\bibfnamefont {S.}~\bibnamefont
  {Gopalakrishnan}}, \bibinfo {author} {\bibfnamefont {K.}~\bibnamefont
  {Agarwal}}, \bibinfo {author} {\bibfnamefont {E.~A.}\ \bibnamefont {Demler}},
  \bibinfo {author} {\bibfnamefont {D.~A.}\ \bibnamefont {Huse}}, \ and\
  \bibinfo {author} {\bibfnamefont {M.}~\bibnamefont {Knap}},\ }\bibfield
  {title} {\enquote {\bibinfo {title} {Griffiths effects and slow dynamics in
  nearly many-body localized systems},}\ }\href {\doibase
  10.1103/PhysRevB.93.134206} {\bibfield  {journal} {\bibinfo  {journal} {Phys.
  Rev. B}\ }\textbf {\bibinfo {volume} {93}},\ \bibinfo {pages} {134206}
  (\bibinfo {year} {2016})}\BibitemShut {NoStop}%
\bibitem [{\citenamefont {Zhang}\ \emph {et~al.}(2016)\citenamefont {Zhang},
  \citenamefont {Zhao}, \citenamefont {Devakul},\ and\ \citenamefont
  {Huse}}]{zhang2016many}%
  \BibitemOpen
  \bibfield  {author} {\bibinfo {author} {\bibfnamefont {L.}~\bibnamefont
  {Zhang}}, \bibinfo {author} {\bibfnamefont {B.}~\bibnamefont {Zhao}},
  \bibinfo {author} {\bibfnamefont {T.}~\bibnamefont {Devakul}}, \ and\
  \bibinfo {author} {\bibfnamefont {D.~A.}\ \bibnamefont {Huse}},\ }\bibfield
  {title} {\enquote {\bibinfo {title} {Many-body localization phase transition:
  A simplified strong-randomness approximate renormalization group},}\ }\href
  {\doibase 10.1103/PhysRevB.93.224201} {\bibfield  {journal} {\bibinfo
  {journal} {Phys. Rev. B}\ }\textbf {\bibinfo {volume} {93}},\ \bibinfo
  {pages} {224201} (\bibinfo {year} {2016})}\BibitemShut {NoStop}%
\bibitem [{\citenamefont {Khemani}\ \emph {et~al.}(2017)\citenamefont
  {Khemani}, \citenamefont {Lim}, \citenamefont {Sheng},\ and\ \citenamefont
  {Huse}}]{khemani2017critical}%
  \BibitemOpen
  \bibfield  {author} {\bibinfo {author} {\bibfnamefont {V.}~\bibnamefont
  {Khemani}}, \bibinfo {author} {\bibfnamefont {S.~P.}\ \bibnamefont {Lim}},
  \bibinfo {author} {\bibfnamefont {D.~N.}\ \bibnamefont {Sheng}}, \ and\
  \bibinfo {author} {\bibfnamefont {D.~A.}\ \bibnamefont {Huse}},\ }\bibfield
  {title} {\enquote {\bibinfo {title} {Critical properties of the many-body
  localization transition},}\ }\href {\doibase 10.1103/PhysRevX.7.021013}
  {\bibfield  {journal} {\bibinfo  {journal} {Phys. Rev. X}\ }\textbf {\bibinfo
  {volume} {7}},\ \bibinfo {pages} {021013} (\bibinfo {year}
  {2017})}\BibitemShut {NoStop}%
\bibitem [{\citenamefont {Luitz}\ and\ \citenamefont
  {Bar~Lev}(2016)}]{luitz2016anomalous}%
  \BibitemOpen
  \bibfield  {author} {\bibinfo {author} {\bibfnamefont {D.~J.}\ \bibnamefont
  {Luitz}}\ and\ \bibinfo {author} {\bibfnamefont {Y.}~\bibnamefont
  {Bar~Lev}},\ }\bibfield  {title} {\enquote {\bibinfo {title} {Anomalous
  thermalization in ergodic systems},}\ }\href {\doibase
  10.1103/PhysRevLett.117.170404} {\bibfield  {journal} {\bibinfo  {journal}
  {Phys. Rev. Lett.}\ }\textbf {\bibinfo {volume} {117}},\ \bibinfo {pages}
  {170404} (\bibinfo {year} {2016})}\BibitemShut {NoStop}%
\bibitem [{\citenamefont {Luitz}\ \emph {et~al.}(2016)\citenamefont {Luitz},
  \citenamefont {Laflorencie},\ and\ \citenamefont {Alet}}]{luitz2016extended}%
  \BibitemOpen
  \bibfield  {author} {\bibinfo {author} {\bibfnamefont {D.~J.}\ \bibnamefont
  {Luitz}}, \bibinfo {author} {\bibfnamefont {N.}~\bibnamefont {Laflorencie}},
  \ and\ \bibinfo {author} {\bibfnamefont {F.}~\bibnamefont {Alet}},\
  }\bibfield  {title} {\enquote {\bibinfo {title} {Extended slow dynamical
  regime close to the many-body localization transition},}\ }\href {\doibase
  10.1103/PhysRevB.93.060201} {\bibfield  {journal} {\bibinfo  {journal} {Phys.
  Rev. B}\ }\textbf {\bibinfo {volume} {93}},\ \bibinfo {pages} {060201}
  (\bibinfo {year} {2016})}\BibitemShut {NoStop}%
\bibitem [{\citenamefont {Luitz}\ and\ \citenamefont
  {Bar~Lev}(2017)}]{luitz2017ergodic}%
  \BibitemOpen
  \bibfield  {author} {\bibinfo {author} {\bibfnamefont {D.~J.}\ \bibnamefont
  {Luitz}}\ and\ \bibinfo {author} {\bibfnamefont {Y.}~\bibnamefont
  {Bar~Lev}},\ }\bibfield  {title} {\enquote {\bibinfo {title} {The ergodic
  side of the many-body localization transition},}\ }\href
  {https://onlinelibrary.wiley.com/doi/abs/10.1002/andp.201600350} {\bibfield
  {journal} {\bibinfo  {journal} {Annalen der Physik}\ }\textbf {\bibinfo
  {volume} {529}},\ \bibinfo {pages} {1600350} (\bibinfo {year}
  {2017})}\BibitemShut {NoStop}%
\bibitem [{\citenamefont {De~Roeck}\ \emph {et~al.}(2016)\citenamefont
  {De~Roeck}, \citenamefont {Huveneers}, \citenamefont {M\"uller},\ and\
  \citenamefont {Schiulaz}}]{deroeck2016absence}%
  \BibitemOpen
  \bibfield  {author} {\bibinfo {author} {\bibfnamefont {W.}~\bibnamefont
  {De~Roeck}}, \bibinfo {author} {\bibfnamefont {F.}~\bibnamefont {Huveneers}},
  \bibinfo {author} {\bibfnamefont {M.}~\bibnamefont {M\"uller}}, \ and\
  \bibinfo {author} {\bibfnamefont {M.}~\bibnamefont {Schiulaz}},\ }\bibfield
  {title} {\enquote {\bibinfo {title} {Absence of many-body mobility edges},}\
  }\href {\doibase 10.1103/PhysRevB.93.014203} {\bibfield  {journal} {\bibinfo
  {journal} {Phys. Rev. B}\ }\textbf {\bibinfo {volume} {93}},\ \bibinfo
  {pages} {014203} (\bibinfo {year} {2016})}\BibitemShut {NoStop}%
\bibitem [{Note1()}]{Note1}%
  \BibitemOpen
  \bibinfo {note} {The absence of energy resolution is most simply understood
  by noting that the distribution of cluster sizes is the same when statistics
  are taken over all clusters from a disorder realisation (which thus includes
  the entire Fock-space site energy distribution), and when they are taken over
  only the clusters $\protect \mathcal {C}$ `grown' from sites with their
  site-energies close to the mean of the Fock-space site energy
  distribution.}\BibitemShut {Stop}%
\bibitem [{\citenamefont {Lazarides}\ \emph {et~al.}(2015)\citenamefont
  {Lazarides}, \citenamefont {Das},\ and\ \citenamefont
  {Moessner}}]{lazarides2015fate}%
  \BibitemOpen
  \bibfield  {author} {\bibinfo {author} {\bibfnamefont {A.}~\bibnamefont
  {Lazarides}}, \bibinfo {author} {\bibfnamefont {A.}~\bibnamefont {Das}}, \
  and\ \bibinfo {author} {\bibfnamefont {R.}~\bibnamefont {Moessner}},\
  }\bibfield  {title} {\enquote {\bibinfo {title} {Fate of many-body
  localization under periodic driving},}\ }\href {\doibase
  10.1103/PhysRevLett.115.030402} {\bibfield  {journal} {\bibinfo  {journal}
  {Phys. Rev. Lett.}\ }\textbf {\bibinfo {volume} {115}},\ \bibinfo {pages}
  {030402} (\bibinfo {year} {2015})}\BibitemShut {NoStop}%
\bibitem [{\citenamefont {Ponte}\ \emph
  {et~al.}(2015{\natexlab{a}})\citenamefont {Ponte}, \citenamefont
  {Papi\ifmmode~\acute{c}\else \'{c}\fi{}}, \citenamefont {Huveneers},\ and\
  \citenamefont {Abanin}}]{ponte2015many}%
  \BibitemOpen
  \bibfield  {author} {\bibinfo {author} {\bibfnamefont {P.}~\bibnamefont
  {Ponte}}, \bibinfo {author} {\bibfnamefont {Z.}~\bibnamefont
  {Papi\ifmmode~\acute{c}\else \'{c}\fi{}}}, \bibinfo {author} {\bibfnamefont
  {F.}~\bibnamefont {Huveneers}}, \ and\ \bibinfo {author} {\bibfnamefont
  {D.~A.}\ \bibnamefont {Abanin}},\ }\bibfield  {title} {\enquote {\bibinfo
  {title} {Many-body localization in periodically driven systems},}\ }\href
  {\doibase 10.1103/PhysRevLett.114.140401} {\bibfield  {journal} {\bibinfo
  {journal} {Phys. Rev. Lett.}\ }\textbf {\bibinfo {volume} {114}},\ \bibinfo
  {pages} {140401} (\bibinfo {year} {2015}{\natexlab{a}})}\BibitemShut
  {NoStop}%
\bibitem [{\citenamefont {Ponte}\ \emph
  {et~al.}(2015{\natexlab{b}})\citenamefont {Ponte}, \citenamefont {Chandran},
  \citenamefont {Papi{\'c}},\ and\ \citenamefont
  {Abanin}}]{ponte2015periodically}%
  \BibitemOpen
  \bibfield  {author} {\bibinfo {author} {\bibfnamefont {P.}~\bibnamefont
  {Ponte}}, \bibinfo {author} {\bibfnamefont {A.}~\bibnamefont {Chandran}},
  \bibinfo {author} {\bibfnamefont {Z.}~\bibnamefont {Papi{\'c}}}, \ and\
  \bibinfo {author} {\bibfnamefont {D.~A.}\ \bibnamefont {Abanin}},\ }\bibfield
   {title} {\enquote {\bibinfo {title} {Periodically driven ergodic and
  many-body localized quantum systems},}\ }\href
  {https://www.sciencedirect.com/science/article/pii/S0003491614003212}
  {\bibfield  {journal} {\bibinfo  {journal} {Annals of Physics}\ }\textbf
  {\bibinfo {volume} {353}},\ \bibinfo {pages} {196--204} (\bibinfo {year}
  {2015}{\natexlab{b}})}\BibitemShut {NoStop}%
\bibitem [{\citenamefont {Roy}\ \emph {et~al.}(2018)\citenamefont {Roy},
  \citenamefont {Bar~Lev},\ and\ \citenamefont {Luitz}}]{roy2018anomalous}%
  \BibitemOpen
  \bibfield  {author} {\bibinfo {author} {\bibfnamefont {S.}~\bibnamefont
  {Roy}}, \bibinfo {author} {\bibfnamefont {Y.}~\bibnamefont {Bar~Lev}}, \ and\
  \bibinfo {author} {\bibfnamefont {D.~J.}\ \bibnamefont {Luitz}},\ }\bibfield
  {title} {\enquote {\bibinfo {title} {Anomalous thermalization and transport
  in disordered interacting floquet systems},}\ }\href {\doibase
  10.1103/PhysRevB.98.060201} {\bibfield  {journal} {\bibinfo  {journal} {Phys.
  Rev. B}\ }\textbf {\bibinfo {volume} {98}},\ \bibinfo {pages} {060201}
  (\bibinfo {year} {2018})}\BibitemShut {NoStop}%
\end{thebibliography}%
\end{document}